\documentclass[iop]{emulateapj}
\usepackage{graphicx}
\usepackage{natbib}
\usepackage{amsmath}
\usepackage{longtable}

\slugcomment{}
\received{2014 May 16}

\shortauthors{Foster \& Brunt}
\shorttitle{A CGPS Look at Outer Galactic Structure}

\begin{document}


\title{A CGPS Look at the Spiral Structure of the Outer Milky Way I:
Distances and Velocities to Star Forming Regions}


\author{Tyler J. Foster\altaffilmark{1,2}}
\affil{Department of Physics \& Astronomy,
Brandon University, Brandon, Manitoba, Canada and\newline
National Research Council Canada, Penticton, British Columbia, Canada}
\email{FosterT@BrandonU.CA}
\and
\author{Christopher M. Brunt\altaffilmark{3}}
\affil{Astrophysics Group, School of Physics, University of Exeter, 
Exeter, United Kingdom}
\email{brunt@astro.ex.ac.uk}

\altaffiltext{1}{Department of Physics \& Astronomy,
Brandon University, 270-18th Street, Brandon, MB, R7A 6A9 Canada}
\altaffiltext{2}{National Research Council of Canada, Emerging Technologies 
-- National Science Infrastructure, Dominion Radio Astrophysical Observatory,
P.O. Box 248, Penticton BC, V2A 6J9, Canada}
\altaffiltext{3}{Astrophysics Group, School of Physics, University of Exeter,
Stocker Road, Exeter, EX4 4QL, United Kingdom}


\begin{abstract}
We present a new catalogue of spectrophotometric distances and line-of-sight 
systemic velocities to 103 \ion{H}{2} regions between 
90$\degr\leq \ell \leq~$195$\degr$ (longitude quadrants II and part of III). 
Two new velocities for each region are independently measured using 1-arcminute 
resolution 21~cm \ion{H}{1} and 2.6~mm $^{12}$CO line maps (from the Canadian
Galactic Plane Survey and FCRAO Outer Galaxy Surveys) that show where gaseous 
shells are observed around the periphery of the ionized gas. Known and 
neighbouring O\&B-type stars with published UBV photometry and MK 
classifications are overlaid onto 21~cm continuum maps, and those stars 
observed within the boundary of the \ion{H}{2} emission (and whose distance is 
not more than 3 times the standard deviation of the others) are used to 
calculate new mean stellar distances to each of the 103 nebulae. Using this 
approach of excluding distance outliers from the mean distance to a group of 
many stars in each \ion{H}{2} region lessens the impact of anomalous reddening 
for certain individuals. The standard deviation of individual stellar distances 
in a cluster is typically 20\% per stellar distance, and the error in the mean 
distance to the cluster is typically $\pm$10\%. Final mean distances of 9 
common objects with VLBI parallax distances show a 1:1 correspondence. Further, 
comparison with previous catalogues of \ion{H}{2} regions in these quadrants 
shows a 50\% reduction in scatter for the distance to Perseus spiral arm 
objects in the same region, and a reduction by $\sim$1/$\sqrt{2}$ in scatter 
around a common angular velocity relative to the Sun 
$\Omega-\Omega_0$(km~s$^{-1}$~kpc$^{-1}$). The purpose of the catalogue is to 
provide a foundation for more detailed large-scale Galactic spiral structure 
and dynamics (rotation curve, density wave streaming) studies in the 
2$^{\textrm{\footnotesize{nd}}}$ and 3$^{\textrm{\footnotesize{rd}}}$ 
quadrants, which from the Sun's location is the most favourably viewed section 
of the Galaxy.
\end{abstract}


\section{Introduction}
The utility of \ion{H}{2} regions as tracers of Galactic spiral structure and 
rotation is well-known. Some of today's most successful models of our Galaxy 
\citep[e.g. the electron-density model of ][and the dust model of 
\citet{drim01}]{tc93} are based on the \ion{H}{2} arms originally mapped by 
\citet{gg76}. Distances to \ion{H}{2} regions and their attendant molecular 
clouds can be obtained photometrically with broadband and spectroscopic data on 
internal exciting stars, and/or kinematically via the object's systemic 
velocity. The classic map of \citet{gg76} uses both methods, as does very 
recent work to produce global maps of spiral structure 
\citep[e.g.][]{hou09,russ03} or more detailed spiral structure maps in smaller 
areas across the Galactic plane \citep[e.g.][]{vazq08}. Neither method can 
approach the accuracy typical of VLBI/parallax observations 
\citep[e.g.][]{xu06} but large-scale mapping efforts with this method are only 
just beginning \citep[see][for a good review, though this method too can be 
quite uncertain; see \citet{miyo12}]{brun11}.


In particular, spiral structure in the 2$^{\textrm{\footnotesize{nd}}}$ and 
3$^{\textrm{\footnotesize{rd}}}$ quadrants of longitude (QII and QIII; 
90$\degr\geq\ell\geq$270~$\degr$), our nearest view of the Outer Galaxy, is
uncertain. Arms are moderately well-defined to about $R\sim$10~kpc \citep[e.g. 
see][]{reid09}, with most maps distinguishing the Local arm (a steeply-pitched 
spur or {}``armlet'' that the Sun seems to be a member of) from the Perseus arm 
(a major tightly-wound star-forming arm, the closest such arm to the Sun). 
Beyond 2-3~kpc from the Sun in QII and QIII, most maps presented thus far 
become quite ill-defined. The modern consensus of spiral structure rendered in 
\citet{chur09} (principally based on \citeauthor{gg76}) shows one arm beyond 
Perseus, the {}``Outer'' arm, some $\sim$6.1~kpc distant from the Sun towards 
$\ell=$180$\degr$, and placed there to kinematically link with the QIII arm of 
\citet{mccl04} in the Outer Galaxy and linking to the Norma arm in the Inner 
Galaxy QIV. In $^{12}$CO $\ell-v$ maps by \citet{dame01} this Outer arm is a 
sparse chain of clouds that appears somewhat coherent over a wide range of 
longitude. \citet{stra07} also trace this arm kinematically using continuum 
absorption towards extragalactic sources. Discrete tracers beyond Perseus in 
this area are also shown in \citet{vazq08} but their fit with the Outer arm 
\citep[as defined by the {}``Cygnus'' arm of][]{vall08} is very poor. While 
these distant tracers suffer from high scatter, if anything their mean position 
seems to be somewhat closer to the Sun than the Cygnus Arm of \citet{vall08}, 
and are more consistent as belonging to the optical arm suggested by 
\citet{kime89} and \citet{negu03}, which at 4-5~kpc from the Sun towards 
$\ell=$180$\degr$ is nearer than the Cygnus arm of \citet{vall08} and the Outer 
arm of \citet{chur09}. This arm is positioned 3.5~kpc towards $\ell=$180$\degr$ 
in the 4-arm model of \citet{hou09} though it is poorly defined and not linked 
to any arms in the Inner Galaxy. Beyond the Cygnus arm, there is unreliable 
evidence for an even more distant Far Outer arm some 7-9~kpc from the Sun 
towards 180$\degr$ \citep[e.g.][]{vall08,hou09}. Such an arm could be a 
QII-extension of the Scutum-Centaurus arm identified in the 
1$^{\footnotesize{\textrm{st}}}$ quadrant by \citet{dame11}. While it is 
proposed by \citet{chur09} and \citet{dame11} that this arm along with Perseus 
forms the principal 2-armed density wave pattern emanating from the central bar 
in the MW, the picture is still not clear \citep[see the synopsis of spiral
structure in][which shows the Perseus arm as a secondary arm]{vall08}.


Clearly then, the view of the overall Milky Way spiral structure and how the
pattern in QII and QIII relates to it is (as yet) incomplete and conflicted. In
particular a more precise location and winding angle for the QII/III arms would
help to mate the patterns and visualize the big picture. The challenge then is 
to refine distances to spiral arm tracers and to minimize their uncertainties, 
with a more rigorous and systematic approach to spectro-photometric distances 
to stars, and/or better defined measurements of the \textit{systemic} 
velocities of tracers. That is the broad purpose of this series of papers.

There are three major obstacles to a better defined map of spiral structure 
from \ion{H}{2} region distances. First and most major are distance 
uncertainties, sources for which are i) misclassification of spectral and MK 
types for exciting stars, ii) errors in photometry, iii) variation in intrinsic 
luminosity for stars of the same class, and iv) a changing reddening law for 
stars that are seen through different relative amounts of interstellar and 
circumstellar dust. Second, data are cobbled together from inhomogeneous sets 
of published catalogues and groups of nebula with different distance estimates 
used for each (as noted, often a mix of kinematic and spectrophotometric), the 
differences among them being constants used (e.g. Solar galactocentric distance 
$R_{0}$), the calibrations of MK/spectral type vs. intrinsic colour and 
magnitudes, and even the identity and the number of stars considered involved 
with a given \ion{H}{2} region. Thirdly, it is not necessary that all observed 
\ion{H}{2} regions must also be tracing a major spiral feature; hence any 
spiral structure present may appear confused by the presence of interarm 
objects and other objects not associated with the major arms. 

Maps made with kinematic distances of \ion{H}{2} regions are also affected by a 
major source of error: deviations of the observed velocity of a tracer from 
circular Galactic rotation. These can come in the form of larger-scale 
systematic non-circular motions like elliptical orbits, and spiral arm dynamics 
\citep[e.g. streaming motions associated with density waves, and {}``rolling'' 
motions that are observed in spiral arms][respectively]{robe72,feit85}, down to 
smaller-scale motions like expansions of gaseous shells surrounding the stars, 
molecular and champagne-type outflows associated with many SFRs, and down to 
the very-small scale random turbulent motions associated with the gas. Even 
such small non-circular deviations can be magnified by the 1/sin$\ell$ 
projection of circular velocities in QII and QIII.

The new catalogue presented here (Paper I) is the foundation for a series of 
following papers that have two purposes; Paper II) clarifies and refines the 
spatial spiral structure in the Outer Galaxy's 2$^{\footnotesize{\textrm{nd}}}$ 
and 3$^{\footnotesize{\textrm{rd}}}$ quadrants, and Paper III) refines the 
observed rotation curve and resolves new detailed spiral arm dynamics within 
it. These goals are accomplished by specifically addressing many of the sources 
of uncertainty noted above. Specifically, these papers improve upon the 
following sources: 1) errors in distances due to random measurement errors 
(such as errors in photometry) and random variations in absolute magnitudes and 
colours among common stellar types, 2) scatter due to inhomogeneity in 
parameters/calibrations used to calculate distances, 3) errors in velocities 
due to expanding shells and outflows along the line-of-sight, and 4) errors in 
velocities due to random turbulent motions. Specifically, the next paper (II) 
will further address 5) scatter in spiral structure maps caused by uncertainty 
in the association of each \ion{H}{2} region with spiral arms or interarm 
regions, and Paper III will address 6), the scatter in the rotation curve 
caused by systematic large-scale {}``rolling'' (i.e. $z$-dependent) motions. 

The new catalogue contains 103 \ion{H}{2} regions observed by the Canadian 
Galactic Plane Survey \citep[CGPS,][]{tayl03} in the region 
90$\degr\leq\ell\leq~$195$\degr$, each with a compilation of associated and 
nearby (on the sky) OB stars with spectral classes and UBV photometric 
measurements, as complete as the current literature to 2013 allows. To minimize 
source 1) above, new group distances for each \ion{H}{2} region are calculated 
using typically more stars than are usually used for each nebula, but only 
those that meet several objective spatial criteria for association. To minimize 
source 2) the same modern magnitude-luminosity and colour index $(B-V)_0$ 
calibrations are used for the whole sample. To address source 3) systemic 
velocities with respect to the LSR are found for each \ion{H}{2} region by   
using high-resolution CGPS line data and looking for velocity channels where 
associated shells and clouds are resolved edge-on with respect to the ionized 
gas, and not face-on. Finally random velocity variations (source 4 above) from 
turbulent and cloud-cloud dispersion motions are minimized by deriving two 
independent velocity estimates, from both \ion{H}{1} and CO maps for each 
\ion{H}{2} region. These efforts result in a more homogeneous, meaningful, and 
most importantly repeatable set of \ion{H}{2} region distances and velocities. 
In Papers II and III that follow this catalogue, we deal with sources of 
uncertainity due to 5) and 6) respectively. 

\section{Observations and Method}
\label{obs}
The cornerstone of our catalog of \ion{H}{2} regions is the new systemic
velocity measurements (with respect to the LSR), which come from 
high-resolution (1 arcminute) $\lambda$21~cm \ion{H}{1} data and 
$\lambda$2.6~mm $^{12}$CO$\left(J=1\rightarrow0\right)$ data. The \ion{H}{1} 
data are entirely from the Canadian Galactic Plane Survey 
\citep[CGPS;][]{tayl03}, whereas CO data are from either CGPS or the Exeter 
FCRAO CO Galactic Plane Survey \citep[described in][]{mott10,brun13}, depending 
on longitude. Our catalog covers \ion{H}{2} regions in the Outer Galaxy only 
($R>R_{0}$) in the longitude range 90$\degr\leq\ell\leq~$193$\degr$ and mainly 
within a latitude of $-$3.5$\degr\leq b\leq$+5.5$\degr$. A high-latitude 
extension was also observed as part of the CGPS 
(99$\fdg$85$\leq\ell\leq~$116$\fdg$96) up to $b=$+17$\fdg$56. The complete CGPS 
dataset of 21~cm line and continuum from 50$\fdg$2$\leq\ell\leq~$193$\fdg$3 and 
$-$3$\fdg$55$\leq b\leq+$5$\fdg$55 are available at the Canadian Astronomy Data 
Centre\footnote{CADC; http://cadc.hia.nrc.ca}.

\subsection{\ion{H}{1} Line Observations}
21~cm \ion{H}{1} line observations used herein were carried out with the
7-element interferometer and 26 metre radio telescopes at the Dominion Radio
Astrophysical Observatory (DRAO) for the CGPS. The final line data product
are 5$\fdg$1$\times$5$\fdg$1 \ion{H}{1} datacubes, which have $\simeq$1 
arcminute resolution in each of 256 channel maps separated by 0.824~km~s$^{-1}$ 
and of 1.32~km~s$^{-1}$ resolution. The line data have a brightness-temperature 
sensitivity of $\Delta T_{B}=$3.5$\sin\delta$~K. More information on the CGPS 
observing and data processing strategy may be found in \citet{tayl03}. 

\subsection{CO Line Observations}
To trace molecular material in the Second Quadrant, we make use of the Five
College Radio Astronomy Observatory (FCRAO) Outer Galaxy Survey 
\citep[OGS;][]{heye98}. The OGS mapped $^{12}$CO J=1$-$0 spectral line emission 
over the longitude range 102$\fdg$5$\leq\ell\leq$~141$\fdg$5 and over latitudes 
$-$3$\fdg$0$\leq$~$b$~$\leq$+5$\fdg$4. For regions outside the OGS coverage, we 
use new FCRAO CO surveys: the Extended Outer Galaxy Survey (E-OGS) extends the 
coverage of the OGS to Galactic longitude $\ell=$193$\degr$, over a latitude 
range $-$3$\fdg$5$\leq$~$b$~$\leq$+5$\fdg$5; and a new survey in the Cygnus 
region connects the OGS to the Galactic Ring Survey \citep[GRS;][]{jack06} 
between longitudes 55$\fdg$7$\leq$~$\ell$~$\leq$102$\fdg$5, over the 
(approximate) latitude range $-$1$\fdg$0$\leq$~$b$~$\leq$+1$\fdg$25. Full 
details of these new surveys are reported elsewhere (Brunt et al. in prep.) 
and are briefly summarized here. All new surveys utilised the 32 pixel SEQUOIA 
focal plane array \citep[][]{eric99} to image the $^{12}$CO and $^{13}$CO 
J=1$-$0 spectral lines at $\sim$45 arcsecond angular resolution in the 
on-the-fly (OTF) mapping mode. The two lines were acquired simultaneously by 
the dual channel correlator (DCC), configured with 1024 channels over a 
bandwidth of 50~MHz at each frequency. The total velocity coverage exceeds 
120~km~s$^{-1}$, and is centered on $v_{LSR}~=-$~40~km~s$^{-1}$ over the range 
of longitudes examined in this paper. The channel spacing is 0.126~km~s$^{-1}$ 
($^{12}$CO) and 0.132~km~s$^{-1}$ ($^{13}$CO); the velocity resolution is 
broader by a factor of 1.21. $^{12}$CO OGS data prepared for inclusion in the 
CGPS were smoothed to 1.319~km~s$^{-1}$, and new E-OGS data are smoothed to 
this resolution as well before analysis in this study. During the survey 
observations, pointing and focus checks were carried out every 3-4 hours, 
shortly after dawn/dusk or after a significant change in source coordinates. 
The data were initially converted to the $T_{A}^{*}$ scale using the standard 
chopper wheel method \citep[][]{kutn81}. We used the OTFTOOL software, written 
by M. Heyer, G. Narayanan, and M. Brewer, to place the spectra on a regular 
22.5$\arcsec$ grid in Galactic $\ell,~b$ coordinates. To achieve this, first 
order baselines were fitted to signal-free regions of each spectrum and the 
root mean square noise amplitude ($\sigma$) of each spectrum was recorded. 
After baseline removal, individual spectra contributing to a single Galactic 
coordinate were assigned a $1/\sigma^{2}$ weighting during the gridding. The 
gridded data were scaled to the radiation temperature scale (T$_{R}^{*}$) by 
dividing by $\eta_{FSS}$~=~0.7 to account for forward scattering and spillover 
losses.

\section{Identifying Exciting Stars}
\label{stardists}
An extensive search of the CGPS 21~cm continuum data was conducted to identify 
known bright \ion{H}{2} regions. We find 103 objects, including most of the
objects in Sharpless' second catalogue \citep[][]{shar59} that fall within
the CGPS survey area, and most of which have known associated stars from a set 
of standard catalogs: \citet{geor70, geor73, cgg78, moff79, cw84, aved84, 
hunt90, glus95, russ07}. In some studies exciting stars have been identified 
from deep observations of a particular nebula \citep[Sh2-138 and Sh2-184, 
in][respectively]{deha99,guet97}, or smaller samples of nebulae 
\citep{lahu85,lahu87}. However, to expand the list of known stars associated 
with each \ion{H}{2} region (especially the larger diffuse ones), we follow a 
systematic procedure.

We begin with 1$\arcmin$-resolution 21~cm continuum emission maps of each
\ion{H}{2} region from the CGPS, contoured (typically 1,3,5~Kelvin levels above 
the background) to clearly delineate the boundaries of the ionized gas. Where 
21~cm emission is not well defined or not detected at all we also use H$\alpha$ 
emission contoured from the composite map of \citet{fink03}. Known exciting 
stars from the standard catalogs above are first overlaid, showing positions 
with respect to the continuum emission. Then, we searched around the centre of 
this emission in SIMBAD for additional OB-type stars not reported in the 
{}``standard'' catalogues. Our main source for additional OB stars in the area 
is the catalogue of \citet{reed03} (and references therein), with some 
additional stellar classes and photometry from smaller catalogues of 
observations of specific stars and Galactic plane regions 
\citep[][]{mart72,cram74,hill77,wram81,mass95,negu03,sota11}. Candidates for 
association were those stars whose position was within or reasonably near to 
the outermost 21~cm continuum or H$\alpha$ emission boundary. Mainly those with 
UBV photometry and published spectral types of O3 to B4 were considered. Where 
they exist we also consider Tycho B-type stars which have no assigned numerical 
sub-type. 

Next, spectroscopic distances to all known and candidate stars are calculated 
from their photometry, spectral types and luminosity classes. Distances for all 
stars use reddenings $E(B-V)=\left(B-V\right)-\left(B-V\right)_{0}$ that are 
calculated from the modern spectral type-intrinsic colour calibration referred 
to in \citet{peca12} (for main sequence stars) and \citet{wegn94} (for 
luminosity classes I-IV). Absolute magnitudes are from the $M_V$ calibration 
compiled by \citet{russ03}. Both of these calibrations are reproduced 
at the end of in Table~\ref{tab1}. A value $R_{V}=~$3.2 for the ratio of 
total-to-differential interstellar extinction is assumed throughout 
\citep[][]{fitz07}. Where multiple values for magnitude $V$, colours 
$B-V,~U-B$, spectral types, and MK Luminosity Classes (LC Ia-V) are listed in 
\citet{reed03} we invariably use the more recent published ones. If no LC is 
given, we use the next most recent published one, and if none at all are known 
we assume the star is LCV (main sequence). If the star has only been classed 
as a B-type star from Tycho photometry, we assume LCV, calculate its reddening 
free index $Q=(U-B)-0.72(B-V)$ and estimate its sub-type (B0-B4) from 
\citet{hend90}. As a last resort in three cases (Sh2-134, 147, 166) where a 
single OB type star with only B and V photometry is present but no temperature 
or luminosity class is known, we assume LCV and estimate the spectral type from 
the ratio of 21~cm radio flux and total IR flux \citep[e.g.][or measured with 
reprocessed IRAS data in the CGPS]{chan95}, a distance-independent method 
originally proposed by \citet{dewd91} that involves estimating the number of 
ionizing photons $N_{rad}$(photons~s$^{-1}$) from the nebula's radio flux 
$S_{\nu}$(Jy) \citep[e.g. see equation p.853][with $T_{e}$ assumed 
$\simeq~$10$^{4}$~K]{hunt90} and the IR luminosity $L_{IR}$($L_{\odot}$) and 
comparing their ratio \citep[which is distance-independent but well calibrated 
with model stellar type; e.g. in][]{ster03}. The method assumes the ionizing 
flux from the embedded star(s) is entirely reprocessed into IR luminosity, and 
results in an upper-limit to the spectral type of the star(s) within.

For a robust mean distance estimate to each nebula we attempt to minimize the 
impact of uncertain stellar distances and unrelated stars. With a set of 355 
stars potentially associated with the 103 nebulae, we use an objective 
procedure of excluding candidates from the mean distance estimate. We consider 
all three dimensions: i) two on the plane of the sky in $\ell,~b$, and ii) the 
third in depth/distance $r$. The first two dimensions are filtered by excluding 
stars on the plane of the sky that are not seen within the boundaries of the 
ionized gas (21~cm and/or H$\alpha$ emission contours), and as well are not 
found inside \ion{H}{1} and/or CO gas shells/walls that surround each nebula.
The third dimension (depth) is screened by excluding stars have conspicuous and 
excessively different distances that are $\sim$3$\sigma$ or more away from the 
mean distance to the other members and candidates. Combinations of both i) and 
ii) are also screened. Examples of excluded stars are shown in 
Figures~\ref{hishells} and \ref{sh173} (see Sec.~\ref{velocities}), marked with 
x-symbols; one star in Sh2-204, and one each in the centres of Sh2-207 and 208 
with $\geq$3$\sigma$ distance differences from the others, one star north of 
Waterloo~1 and four stars around Sh2-173 excluded by their position outside of 
the continuum emission and associated \ion{H}{1} and $^{12}$CO shells, and one 
star in Sh2-168 by both its outlying position and distance. Stars that were 
borderline to meeting i) were additionally scrutinized by their published 
radial velocity (if available); those with a similar velocity to the nebula 
($\lesssim~$20~km~s$^{-1}$ different) were included. Finally, for three nebulae 
(Sh2-193, 203 and 232) only two stars are associated with very different 
distances; we choose the single star whose distance is closest to the group of 
\ion{H}{2} regions that these objects belong to (i.e. Sh2-192 to 196, Sh2-203 
to BFS~31, and Sh2-231 to 235 respectively). Applying the above criteria, we 
cull some 45 stars from our list.

We present our full catalogue of 355 stars found in and around Galactic 
\ion{H}{2} regions online in both Excel sheet and PDF formats; see
ftp://ftp.drao.nrc.ca/pub/users/foster/Table1.xls or /Table1.pdf. A sample of 
Table~\ref{tab1} is reproduced below. The 45 stars excluded by the criteria 
above are marked in Table~\ref{tab1} for clarity, but do not contribute to the 
final distance calculation. The final tally of stars we associate spatially 
with 21~cm continuum from each of 103 \ion{H}{2} regions is 310. 205 have been 
classified as LCV in the literature from spectra, and 44 have been assumed 
(mainly in the literature, or by us) LCV for the distance calculation. Among 
them, 23 B-type stars have spectral subtypes estimated from the $Q$ index; one 
or more stars in the following \ion{H}{2} regions: Sh2-129, CTB~104b, 154, 
BFS~17, 157, 163, 170(2), 173(2), 177(3), 199, 204(2), 207(2), 
208/Waterloo~1(2), 249(2), and 259. The number distribution of spectral types 
among O3-B4V stars is plotted in Figure~\ref{lcvhist}, and shows that the most 
common main sequence star in our sample is type B1. 



\begin{deluxetable*}{ccccccccccccccc}


\tabletypesize{\tiny}

\setlength{\tabcolsep}{0.008in}


\tablecaption{Sample lines from full Table~1 online at
ftp://ftp.drao.nrc.ca/pub/users/foster/Table1.pdf. The online version
also contains star names, coordinates, references, and descriptions.}

\tablenum{1}


\tablehead{\colhead{Object} & \colhead{$\ell$} & \colhead{$b$} &
\colhead{$v_{CO}$} & \colhead{$v_{HI}$} & \colhead{$v_{H\alpha}$} &
\colhead{$\langle r \rangle$} & \colhead{$\Delta D$} & \colhead{E(B--V)} 
& \colhead{V} & \colhead{(B--V)} & \colhead{Spec} & \colhead{LC} &
\colhead{(B--V)$_{0}$} & \colhead{M(R03)} \\
\colhead{} & \colhead{(deg.)} & \colhead{(deg.)} & \colhead{(km~s$^{-1}$)} &
\colhead{(km~s$^{-1}$)} & \colhead{(km~s$^{-1}$)} & \colhead{(kpc)} & \colhead{$/\langle r \rangle$} & \colhead{(mag.)} & \colhead{(mag.)} &
\colhead{mag.} & \colhead{Type} & \colhead{} & \colhead{mag.} & \colhead{mag.}}


\startdata
Sh2-121 & 90.23 & 1.72 & --62.4 & --56.7 & --61.1 & 6.82 & 0.05 & 2.08 &
14.91 & 1.86 & B0 & II & --0.22 & --5.75
\\
 &  &  &  &  &  &  &  & 2.066 & 15.44 & 1.76 & O4 & V & --0.306 & --5.52
\\
 &  &  &  &  &  &  &  & 1.92 & 15.48 & 1.76 & B2 & II & --0.16 & --4.8
\\
\tableline
Sh2-124 & 94.57 & --1.45 & --43.6 & --41.865 & --37 & 3.78 & 0.17 & 1.248
& 12.28 & 0.95 & O7 & V & --0.298 & --4.94
\\
 &  &  &  &  &  &  &  & 1.39 & 14.46 & 1.18 & B2 & V & --0.21 & --2.47
\\
\tableline
CTB104b & 94.72 & --1.54 & --15.0 & --14.65 &  & 2.36 & 0.06 & 0.62 & 11.36
& 0.37 & B1.5 & V & --0.25 & --2.81
\\
 &  &  & --41.0 &  &  &  &  & 0.3 & 9.97 & 0.09 & B2 & V & --0.21 & --2.47
\\
 &  &  &  &  &  &  &  & 0.82 & 10.73 & 0.52 & B0 & V & --0.3 & --3.9
\\
 &  &  &  &  &  &  &  & 0.81 & 9.62 & 0.51 & O7.5 & V & --0.3 & --4.88
\\
 &  &  &  &  &  &  &  & 0.41 & 9.06 & 0.11 & B0 & V & --0.3 & --3.9
\\
\tableline
Sh2-127 & 96.27 & 2.57 & --94.7 & --92.98 & --98.9 & 9.97 &  & 1.73 & 15.80
& 1.44 & O8 & V & --0.29 & --4.73
\\
\tableline
BFS8 & 96.35 & --0.2 & --57.5 & --55.47 &  & 8.75 &  & 1.338 & 13.58 & 1.035
& O5 & V & --0.303 & --5.41
\\
\tableline
Sh2-128 & 97.56 & 3.16 & --74.0 & --70.72 & --76.8 & 8.06 &  & 1.768 & 15.25
& 1.47 & O7 & V & --0.298 & --4.94
\\
\tableline
Sh2-129 & 99.06 & 7.4 & 1.4 & --6.3 & --  & 0.81 & 0.18 & 0.38 & 6.09 & 0.08
& B0 & V & --0.3 & --3.9
\\
 &  &  &  & Cep OB2 & --8 &  &  & 0.368 & 8.14 & 0.19 & B3 & V & --0.178 &
--1.6
\\
 &  &  &  &  &  &  &  & 0.4 & 9.21 & 0.19 & B2 & V & --0.21 & --2.47
\\
 &  &  &  &  &  &  &  & 1.06 & 11.17 & 0.85 & B2 & V & --0.21 & --2.47
\\
 &  &  &  &  &  &  &  & 0.65 & 8.63 & 0.44 & B2 & V & --0.21 & --2.47
\\
\enddata  



\label{tab1}
\end{deluxetable*}

Table~2 in this paper gives the final heliocentric stellar distance 
$r\pm dr$ to each of 103 nebulae in the Outer Galaxy. We also include mean 
stellar distances calculated to seven additional \ion{H}{2} regions (associated 
with 10 additional stars) just outside the CGPS high-longitude border (Sh2-261, 
and Sh2-267 through 272). Although these objects could not be scrutinized in 
the same way as regions observed in the CGPS 21~cm maps and (thus) are not part 
of our main catalogue, Sh2-267 - 272 in particular are an important cluster of 
six \ion{H}{2} regions tracing the extension of the Cygnus spiral arm into the 
3$^{\textrm{\footnotesize{rd}}}$ quadrant (discussed in Paper II), so their 
distances are presented here for use later on.

\subsection{Distance Uncertainties}
\begin{figure}
\begin{center}
\includegraphics[angle=-90,width=8cm]{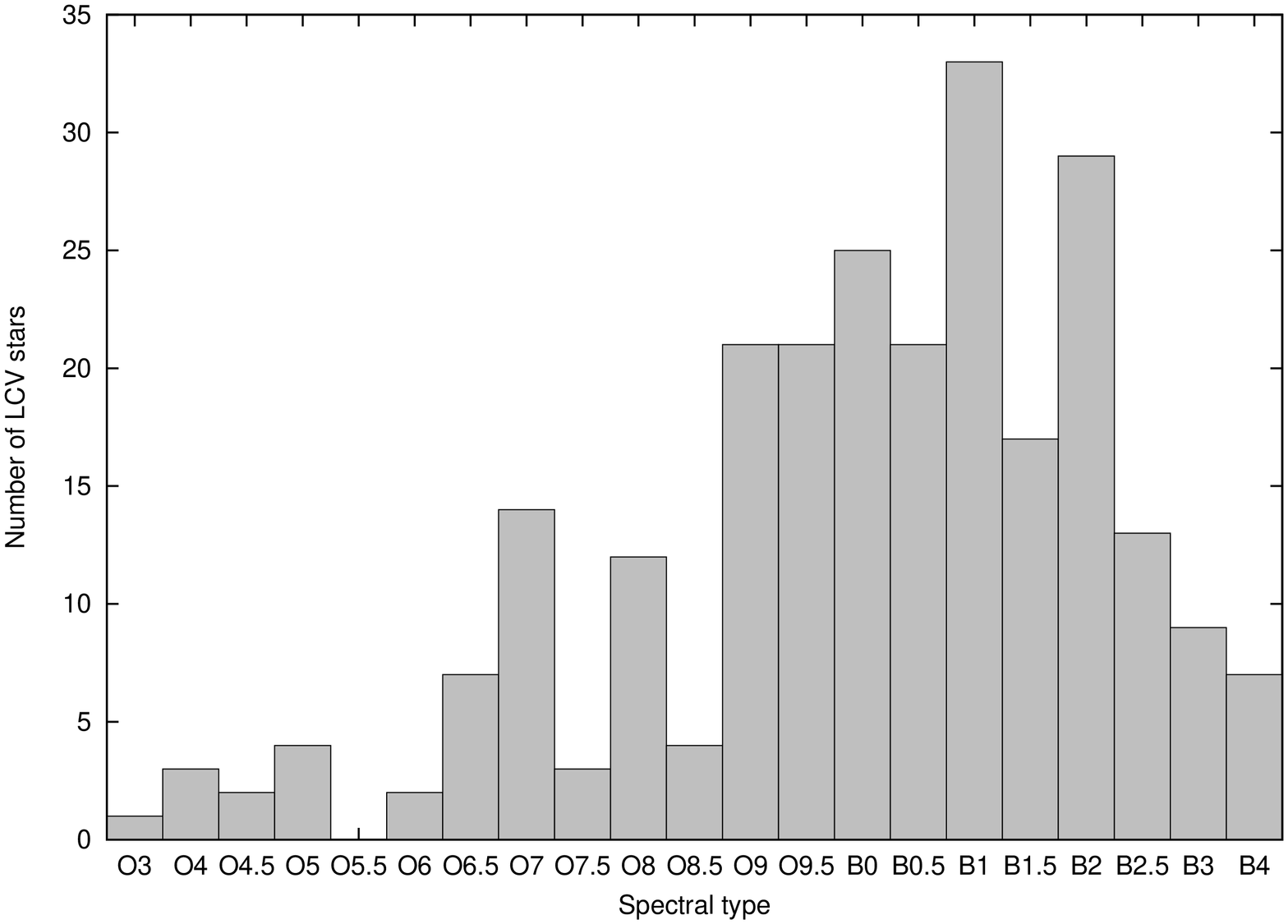}
\includegraphics[angle=-90,width=8cm]{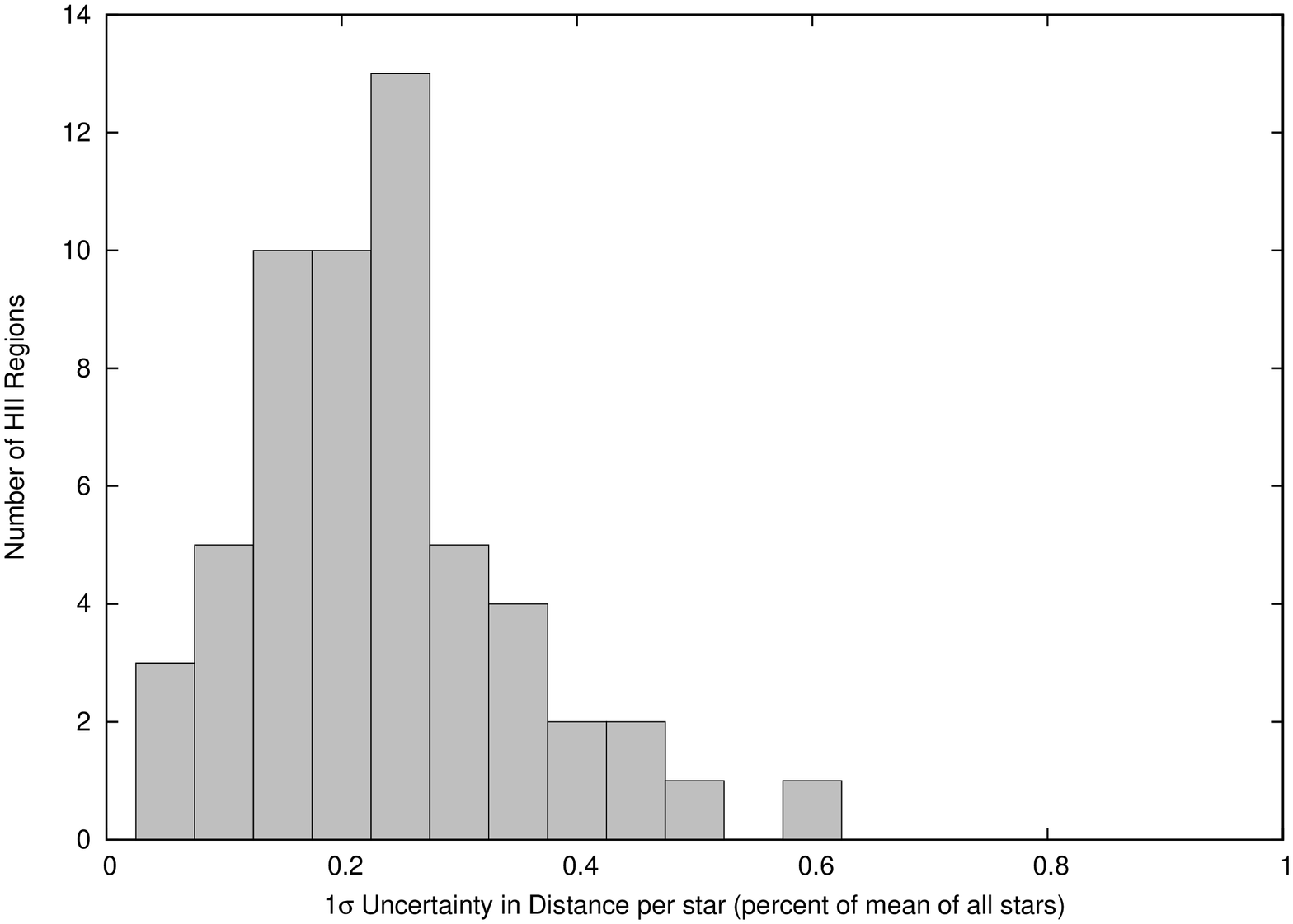}
\includegraphics[angle=-90,width=8cm]{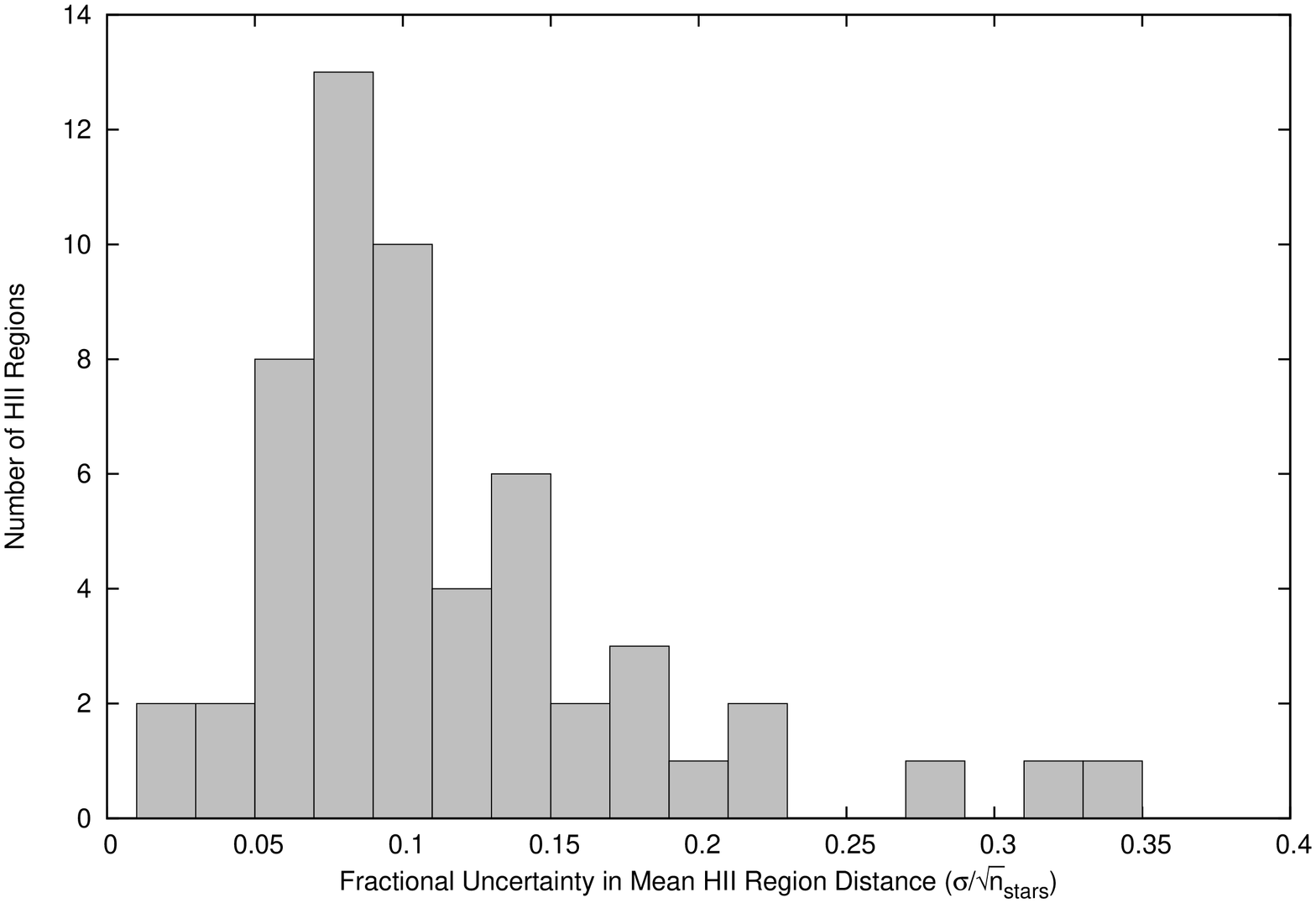}
\end{center}
\caption{(Top) Number distribution of 248 main-sequence type (LCV) stars in our 
main catalogue of 103 \ion{H}{2} regions, with spectral types O3-B4. (Middle) 
Distribution of 56 fractional uncertainties in the distance per star in a 
cluster belonging to an \ion{H}{2} region, calculated from the standard 
deviation of two or more individual stellar distances for the region 
($\sigma_*/r$). (Bottom) Distribution of 56 fractional uncertainties in the 
mean distances to the \ion{H}{2} region's cluster, $\sigma_*/\sqrt{n_*}/r$.}
\label{lcvhist}
\end{figure}

56 of the nebulae (including two groups at a common distance: Sh2-156+BFS~17, 
and Sh2-254-258, and three non-CGPS \ion{H}{2} regions beyond 
$\ell=~$193$\degr$) have two or more associated stars identified (Column 8
in Table~2; 
numbers in brackets are the number of excluded stars), so uncertainties in the 
mean distance to the \ion{H}{2} region can be assessed directly with the 
standard deviation of stellar distances in a cluster of $n_*$ stars as 
$\sigma_*/\sqrt{n_*}$. Here $\sigma_*$ is the standard deviation of the stars' 
distances, or the average uncertainty in an individual star's distance. 
The distributions of distance uncertainties (as a percent of the mean distance 
to the cluster), both per star and for the mean distance to the group are shown 
in Figure \ref{lcvhist}. The errors in the mean distances fairly follow a 
normal distribution with a small positive skew, peaking at $\sim\pm$10\% of the 
mean distance, and range from $\leq$2\% to 30\%. Note that the errors defining 
these distributions include contributions from random errors in photometry and 
errors in spectral classification (temperature and luminosity classes), but not 
from systematic differences caused by, for example, an anomalous extinction law 
(i.e. $R_{V}=A_{V}/E(B-V)>$3.2 for dust of different grain size than that 
typical of the ISM), which is not a normally-distributed error. This will cause 
additional scatter towards higher distances and potentially create some 
dramatically outlying distances (dealt with by robust statistics in Paper II). 
All other nebulae have only one identified stellar member and their distances 
are assigned $\sim\pm$20\%, the average standard deviation per star in all 
groups of two or more stars.

\section{Estimating CO and \ion{H}{1} Velocities from CGPS Data}
\label{velocities}
\begin{figure}
\begin{center}
\vspace{-2.4cm}
\includegraphics[angle=0,width=9cm]{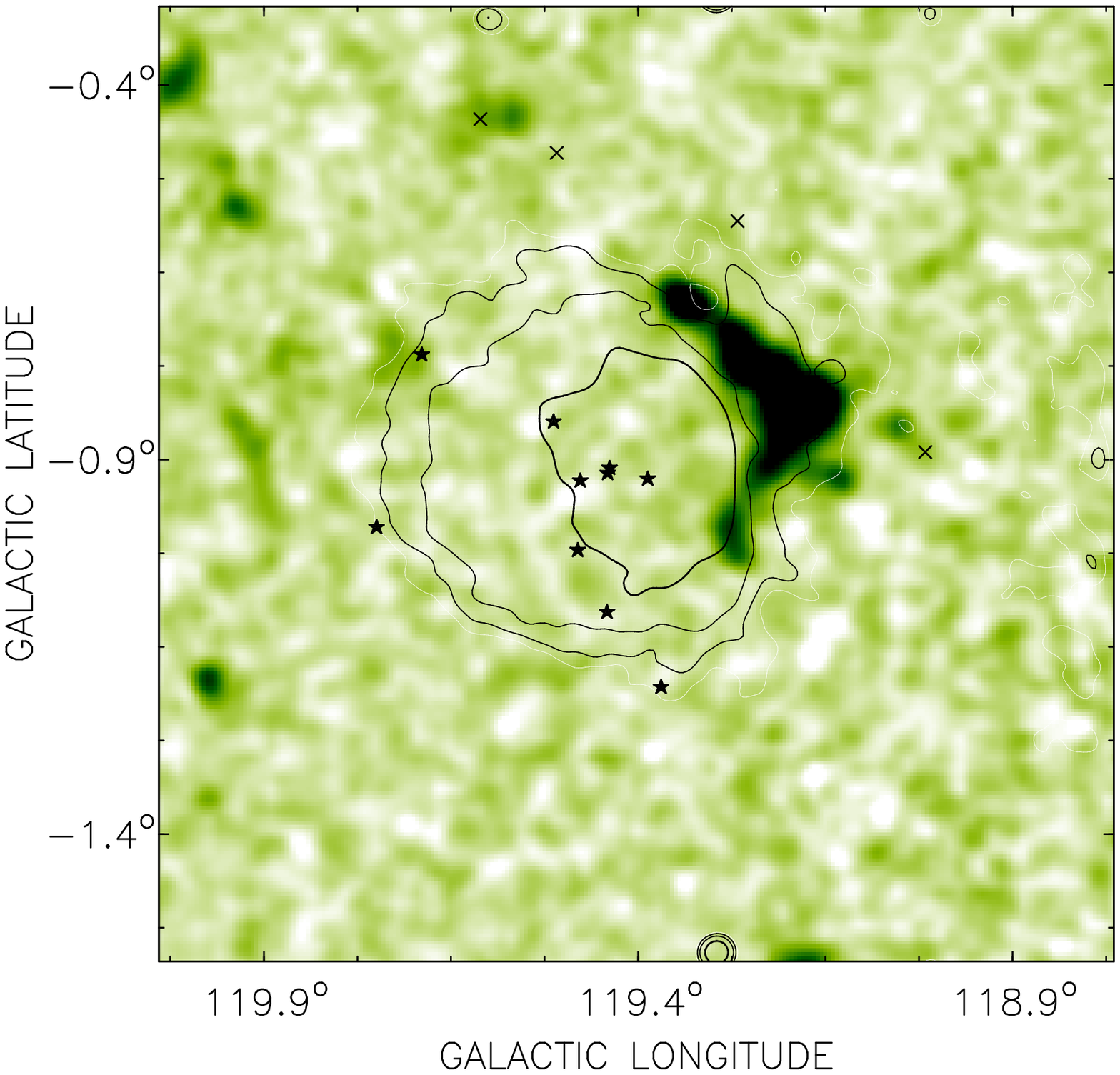}\vspace{-5.2cm}
\includegraphics[angle=0,width=9cm]{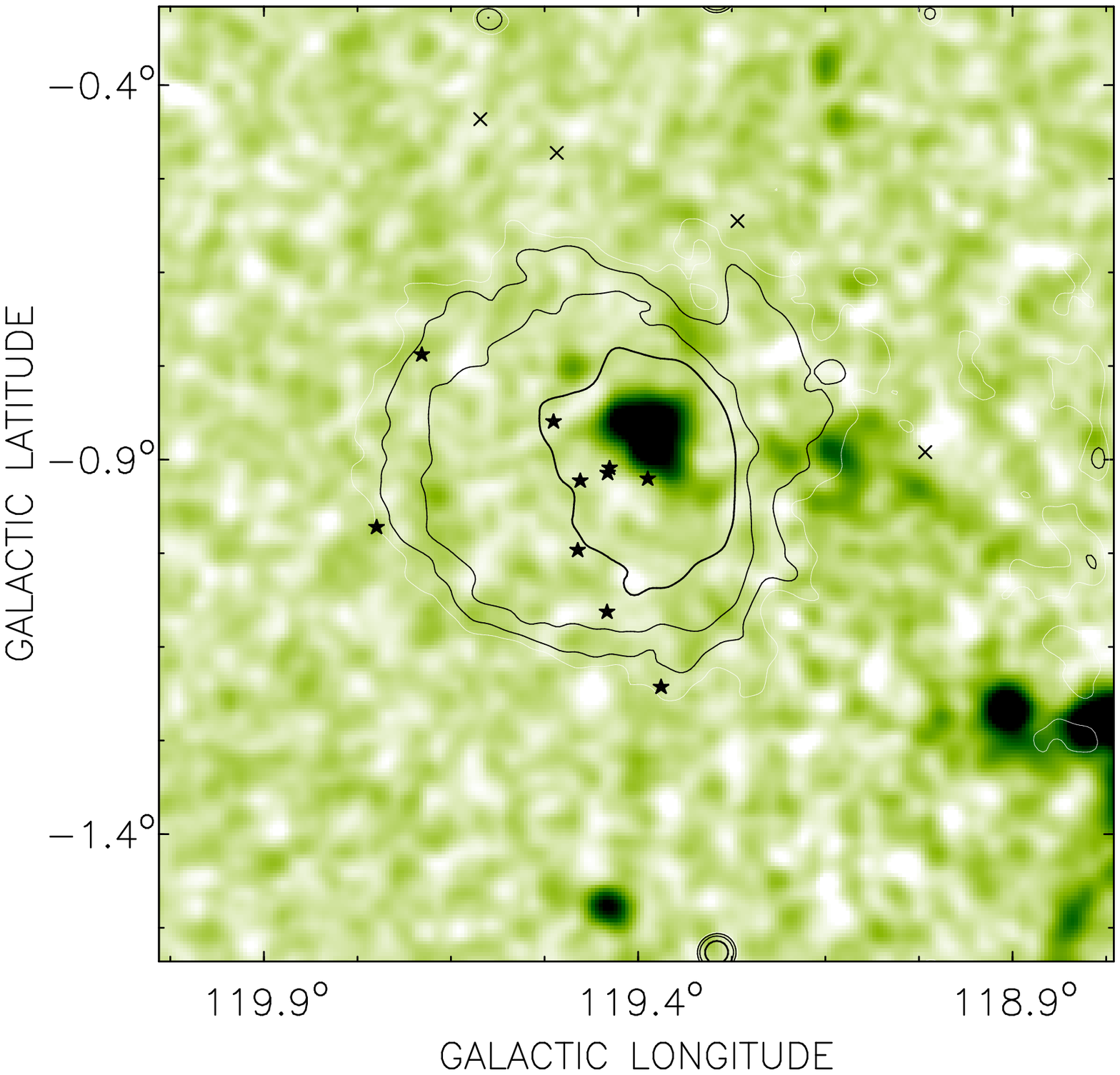}\vspace{-1.6cm}
\end{center}
\caption{Two velocity channels of $^{12}$CO associated with Sh2-173, with 21~cm 
continuum contours (5.4, 5.45, 5.6, 6~K) overlaid, and the positions of 14 
stars associated with the nebula (marked with star symbols), four of which are 
excluded under criteria i) discussed in Sec.~\ref{stardists} (x-symbols). (Top) 
A partial $^{12}$CO shell seen on the edge of the \ion{H}{2} at 
$v_{LSR}=-$30.7~km~s$^{-1}$. (Bottom) $^{12}$CO cloud seen on the face at 
$v_{LSR}=-$38.2~km~s$^{-1}$. This cloud is more likely to be moving towards or 
away from the Sun as part of an expanding shell, making its velocity as an 
indicator of Galactic rotation less reliable than the shell seen edge-on.
}
\label{sh173}
\end{figure}

A telescope pointed directly towards an \ion{H}{2} region is likely to view 
line-of-sight (LOS) molecular outflows, or a molecular shell's front or back 
{}``caps'' that are expanding along the LOS. Typical $^{12}$CO velocities as 
published \citep[e.g.][]{blit82} are almost certainly affected, and thus there 
may be a significant difference between the object's published velocity and the 
\textit{systemic}, which we define as free of motions from LOS expansions and 
outflows, and other systematic effects (e.g. {}``rolling'' motions in the 
arms). The 1/sin$~\ell$ projection from $v_{LSR}$ to angular velocity $\Omega$ 
will amplify any such differences. For Galactic structure and dynamics studies, 
the ideal is to isolate only the LOS component of an object's orbital motion. 
CGPS data are advantageous in this regard: the large-scale high-resolution maps 
allow one to look for the velocity channels where the associated gas is seen 
\textit{to one side of} or \textit{around the edges of} the ionized gas, and 
discriminate against gas seen atop the face of the \ion{H}{2} (see 
Figure~\ref{sh173} for an example). Velocities derived in this way are more 
likely to reflect only the object's motion about the Galactic centre, and less 
likely to exhibit peculiar motions along the LOS, whether systematic or random 
(e.g. {}``turbulence'', cloud-cloud dispersion motions).

Two systemic velocities are estimated independently from CGPS \ion{H}{1} and
from Exeter FCRAO $^{12}$CO(J$=$1$\rightarrow$0) GPS data \citep[][]{mott10,
brun13} for each \ion{H}{2} region. High-resolution $^{12}$CO data towards 
Sh2-121, CTB~104b, 124, 127, 128, 129, 131, and 184 are not available and CO
velocities are estimated with 8-arcminute $^{12}$CO data from \citet{dame01}. 
Additional \ion{H}{2} regions Sh2-261, and 267--272 are not observed in either
$^{12}$CO or \ion{H}{1}, so only CO velocities from \citet{fbs89} (where 
observed) are reported. For each \ion{H}{2} region we first overlay 21~cm 
continuum contours and mark the position of each star on both $^{12}$CO and 
\ion{H}{1} channel maps, and search each channel for structure in the gas that 
correlates with the spatial appearance of the ionized gas and star locations. 
Channel maps where atomic and molecular gas shells that surround the ionized 
gas (defined with 21~cm continuum images) are seen \textit{edge-on} (and not 
\textit{face-on}) are sought. $^{12}$CO emission is naturally concentrated into 
discrete clouds, and it is usually straightforward to identify associated CO 
around the periphery of an \ion{H}{2} region. The ubiquity of \ion{H}{1} 
appearing at all angular scales ($\sim$1$\arcmin$ and up) and across many 
velocity channels creates a somewhat more confused picture, where shells, walls 
and other features may be superposed with ISM clouds, threads and sheets that 
are rather related to larger scale Galactic structure (e.g. the disk, spiral 
arms). To enhance our ability to identify in particular \ion{H}{1} related to 
each \ion{H}{2} region, we also overlay H$\alpha$ emission contours from the 
Wisconsin H-Alpha Mapper (WHAM) plus Virgina Tech Spectral-line Survey (VTSS) 
map \citep[][]{fink03} for comparison (only where the angular resolution of the 
WHAM+VTSS map is the nominal $\sim$6$\arcmin$). A few \ion{H}{2} regions show 
little distinct 21~cm continuum emission above the background (e.g. Sh2-177, 
BFS~28), so for these we rely solely on H$\alpha$ to delineate the ionized gas 
boundaries. 



\begin{widetext}
\begin{figure*}
\begin{center}
\vspace{-3cm}
\includegraphics[angle=0,width=9cm]{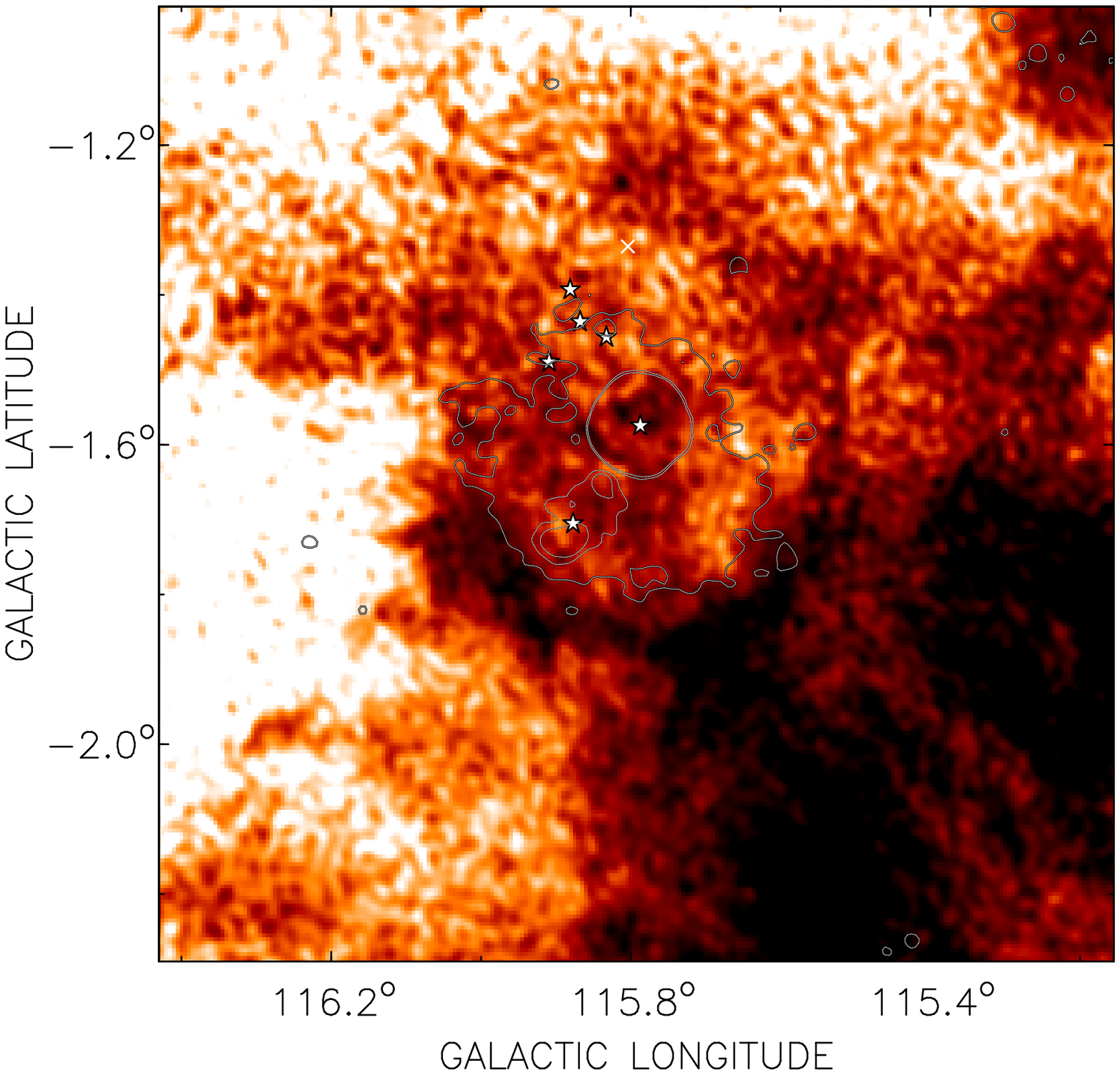}\hspace{-1.3cm}
\includegraphics[angle=0,width=9cm]{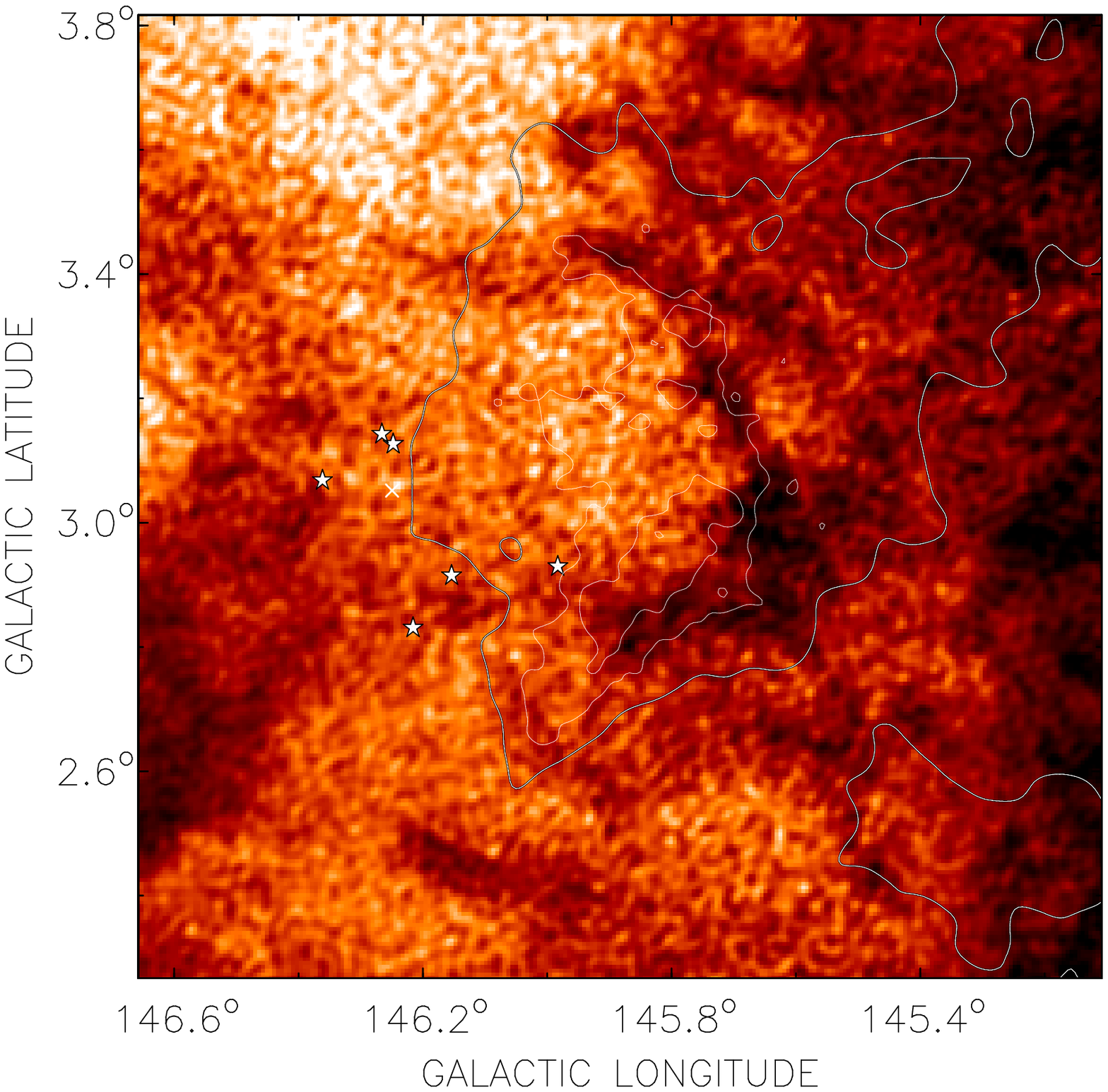}\vspace{-4.2cm}
\includegraphics[angle=0,width=9cm]{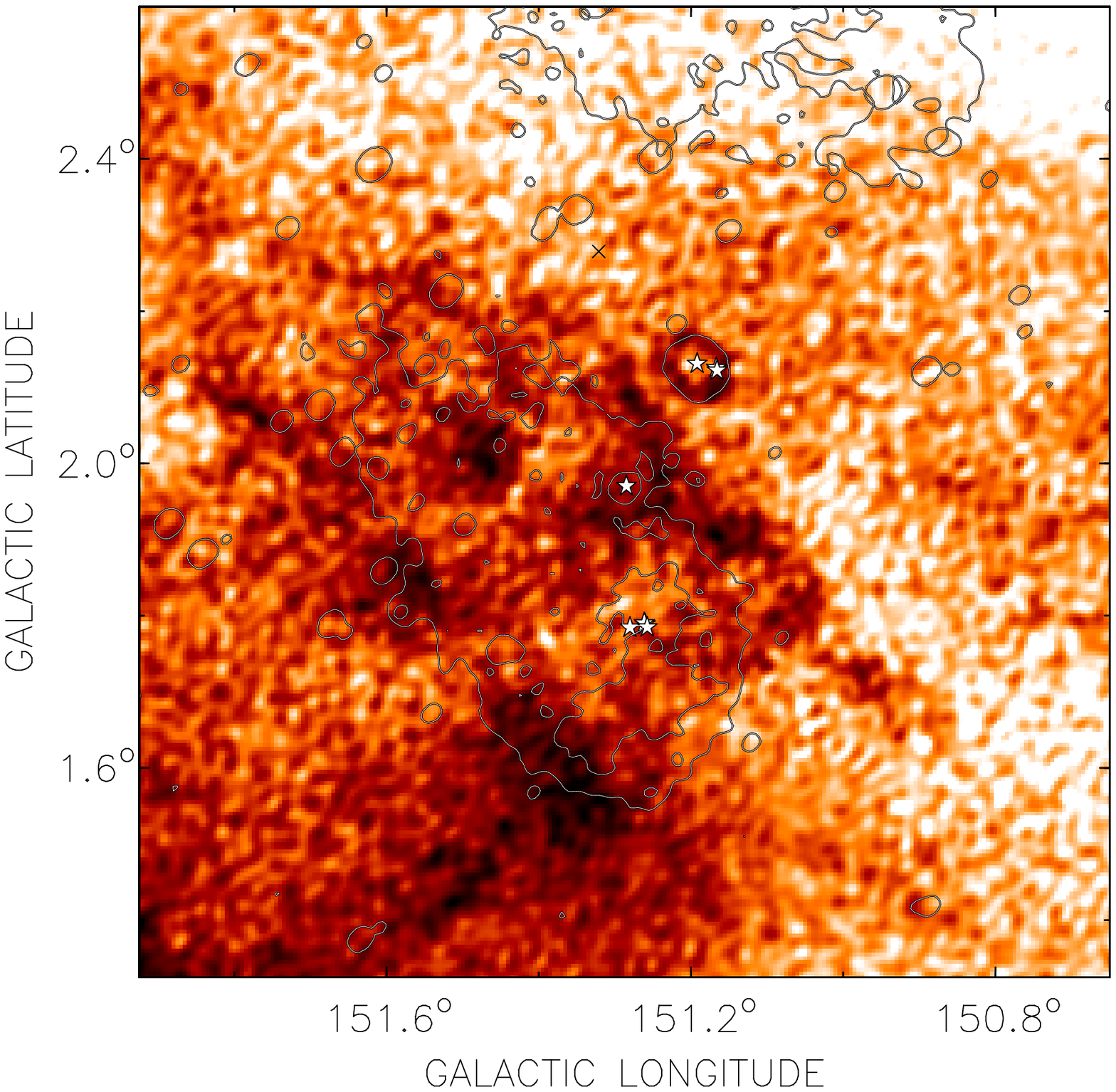}\hspace{-1.3cm}
\includegraphics[angle=0,width=9cm]{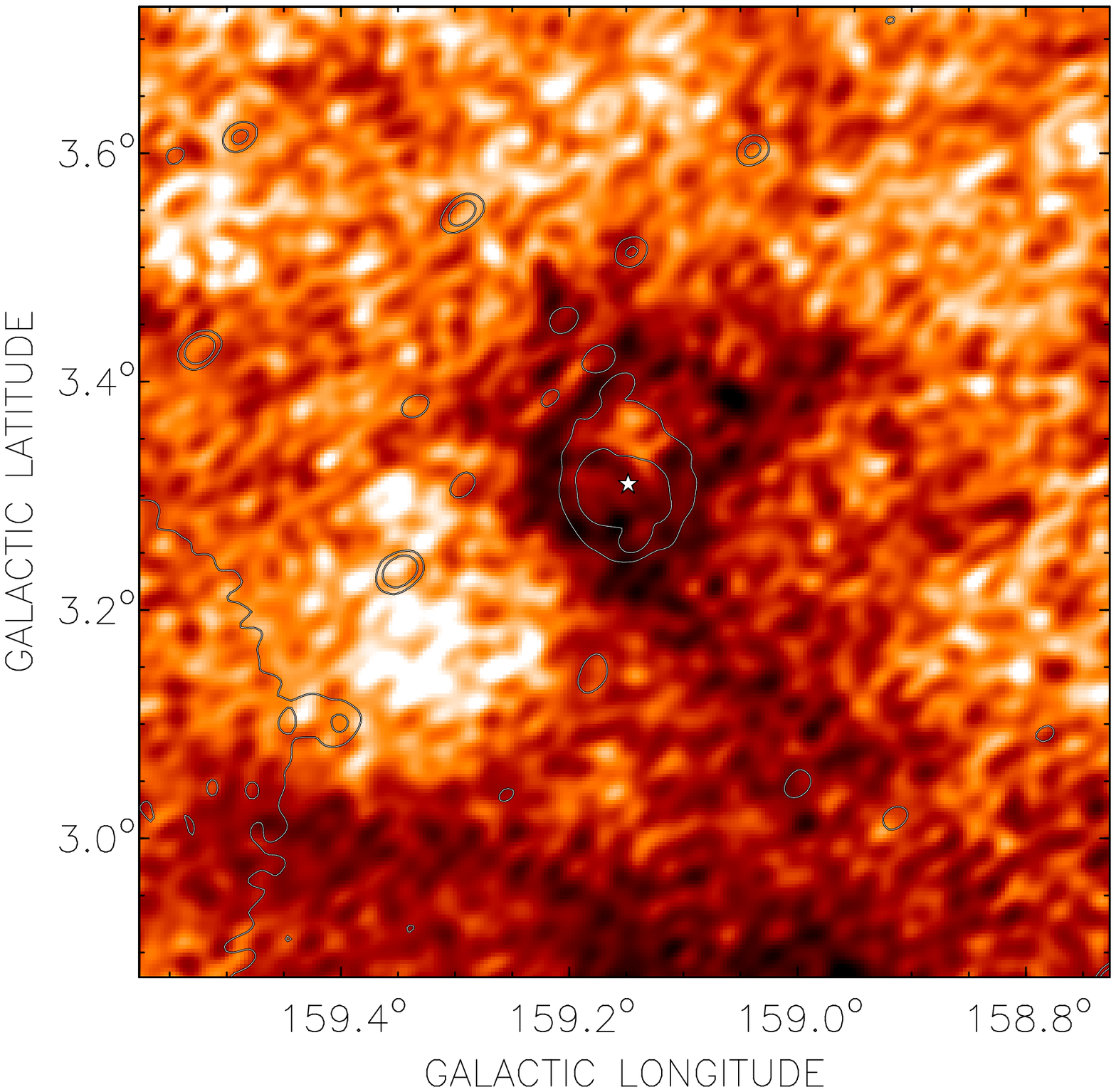}\vspace{-1.5cm}
\end{center}
\caption{Examples of the partial or complete shells of neutral hydrogen 
emission seen for many of the \ion{H}{2} regions in Table~2. Dark 
shading corresponds to bright emission, and the positions of associated stars 
are marked with star symbols. (Top left) A partial semicircular \ion{H}{1} 
shell around the Southern edge of the diffuse emission that surrounds Sh2-168 
\& Sh2-169 (21~cm contours 5.55, 5.8, 5.9~K) at $v_{LSR}=-$34.4~km~s$^{-1}$; 
(Top right) Partial \ion{H}{1} shell surrounding the western edge of Sh2-204 
(21~cm contours 4.9, 5.0, 5.1~K), $v_{LSR}=-$32.4~km~s$^{-1}$. (Bottom left) 
Complete and partial shells surrounding Sh2-207 (top right), and Sh2-208 
(centre) which is embedded in Waterloo~1 (diffuse extended emission) and the 
\ion{H}{1} surroundings at the same velocity $v_{LSR}=-$25.4~km~s$^{-1}$. 
(Bottom right) Complete \ion{H}{1} shell surrounding Sh2-217 (21~cm contours 
5.3 \& 7~K); $v_{LSR}=-$20.4~km~s$^{-1}$.}
\label{hishells}
\end{figure*}
\end{widetext}

In the 1-arcminute data, $^{12}$CO is often seen to form crescent-like shells 
curving partially (e.g. Sh2-173; see Figure \ref{sh173}) or fully (e.g. 
Sh2-231) around the limb of the ionized gas. The systemic velocity is found 
from the centre of several channels where the appearance of the edge-on shell 
is widest and its shape is unchanging from one channel to the next. Often, an 
imprint of the CO features is seen in the \ion{H}{1} channel maps in the form 
of \ion{H}{1} Self-Absorption (HISA; e.g. Sh2-207, Sh2-232) at the same 
velocities, and in general, at the position around the edge where $^{12}$CO is 
seen, corresponding \ion{H}{1} emission from the shell is not. In CGPS 
\ion{H}{1} datacubes we find a wider variety of structures that can be used to 
narrow down the systemic velocity range. These are broadly categorized below
(in order of significance to estimating a systemic velocity):

\begin{enumerate}
\item Thin semi-circular crescents of \ion{H}{1} emission contoured around the 
edges of the ionized gas in a continuous or fragmented arc, and either 
completely surrounding the \ion{H}{2} (e.g. Sh2-207, Sh2-217) or only partially 
so (e.g. Sh2-168-169, Sh2-204; see Figure \ref{hishells}). Annulus or 
crescent-like shells appear completely in a small number of velocity channels. 
The systemic velocity is the mean channel where the shell appears widest and 
unchanging from one channel to the next. 
\item \ion{H}{1} appears to consist of an arc-like portion of a shell contoured 
partway around the \ion{H}{2} boundary. The position of the \ion{H}{1} arc 
around the edge of the \ion{H}{2} moves clockwise or anticlockwise around the 
periphery of the \ion{H}{2} from channel to channel. As a result, the 
velocity-integrated \ion{H}{1} emission forms a partial or complete shell 
around the \ion{H}{2} (e.g. Sh2-139, 198, 254). 
\item Thick flat {}``walls'' of atomic hydrogen emission with a slight 
concavity, inside which the \ion{H}{2} region appears with one or more stars 
whose wind(s) probably are shaping one side of the \ion{H}{1} wall (e.g. 
Sh2-124, 143, 164, 202, 227, 231, 249), and producing an \ion{H}{2} 
{}``blister'' off of the \ion{H}{1} wall.
\item A clear cavity or depression in \ion{H}{1} brightness containing the 
star(s) (e.g. Sh2-141, Sh2-168) and the \ion{H}{2} region(s), perhaps bounded 
on one side by a thick \ion{H}{1} wall or cloud (e.g. Sh2-161, 177, 198, 
LBN~676, Sh2-232) or by a thin nearly complete shell of \ion{H}{1} emission
(e.g. the group Sh2-147 \& 148/149).
\item Small HISA features off the limb of the \ion{H}{2} (e.g. Sh2-192 \& 193, 
205, 206, 228, 234, 252), indicating the presence of a cool compressed edge of 
\ion{H}{1}, or on the face of the \ion{H}{2} (e.g. Sh2-242), indicating a shell 
end-cap. 
\item \ion{H}{1} Continuum Absorption (HICA) of compact unresolved \ion{H}{2} 
regions (e.g. Sh2-121, 138, 156, 211, 255, 258), possibly by shells of dense 
neutral hydrogen expanding outwards from them \citep[e.g.][]{koth02}.
\end{enumerate}

Several examples of the above \ion{H}{1} structures are shown in 
Figure~\ref{hishells}. Instead of presenting the more than 200 such 
figures for each \ion{H}{2} region, for reproducibility we offer detailed text 
descriptions of the observed spatial and velocity structures channel-by-channel 
for each systemic velocity reported in Table~\ref{tab1}. These observing notes 
are viewable in the Excel version of Table~\ref{tab1} as descriptive comments 
attached to each velocity measurement in $^{12}$CO and \ion{H}{1}, in columns 4 
\& 5 respectively. 
This table also includes information to reproduce every observation: continuum 
contour levels to overlay, Galactic coordinates of all 355 stars, as well as 
the compiled list of stellar distance calculations. Readers interested in 
reproducing results with our methodology are encouraged to download CGPS 21~cm 
continuum, $^{12}$CO and \ion{H}{1} line data cubes (see Sec.~\ref{obs}) and 
inspect them along with our descriptions in the online Table1.xls to identify 
an object's molecular and atomic gas morphologies. The final systemic 
velocities in each gaseous tracer for each \ion{H}{2} region in this paper are 
given in Table~2 below.

\subsection{Estimating the Velocity Uncertainty}
Most \ion{H}{2} regions in this catalogue have been scrutinized for related 
\ion{H}{1} and $^{12}$CO at least three times over the 3 year period of this 
study, and where more than one potentially associated feature was identified in 
the datacube we list multiple candidate velocities for each tracer. The 
velocity differences among multiple identified features are typically 
2-5~km~s$^{-1}$ except for a few cases (for example Sh2-151, 166, 168-169, 180, 
G122.6+1.6, G127.1+0.9) where two or more associated \ion{H}{1} and/or CO 
components are found in quite different velocity ranges ($>$8~km~s$^{-1}$ 
different). The above differences reflect the uncertainty inherent in 
estimating velocities by eye: relating structure in channel maps 
to the continuum and H$\alpha$ morphology is a somewhat subjective process. 
The average of these independent estimates then is the final systemic velocity 
measurement reported in Table~2, one for each molecular and atomic 
gas component ($v_{^{12}\textrm{CO}}$ and $v_{\textrm{HI}}$ respectively). 

\begin{figure}
\begin{center}
\includegraphics[angle=-90,width=8cm]{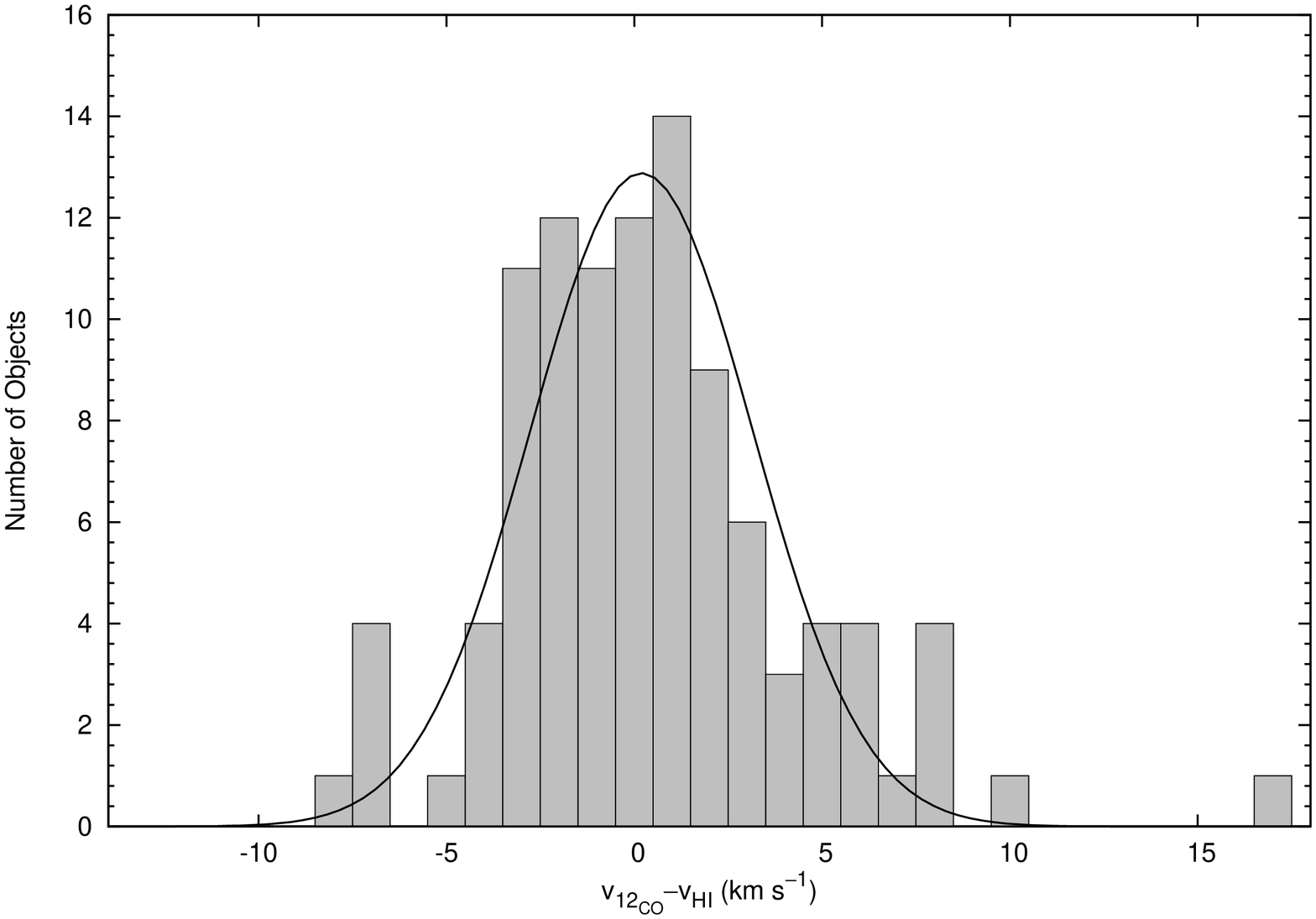}\vspace{-0.25cm}
\includegraphics[angle=-90,width=8cm]{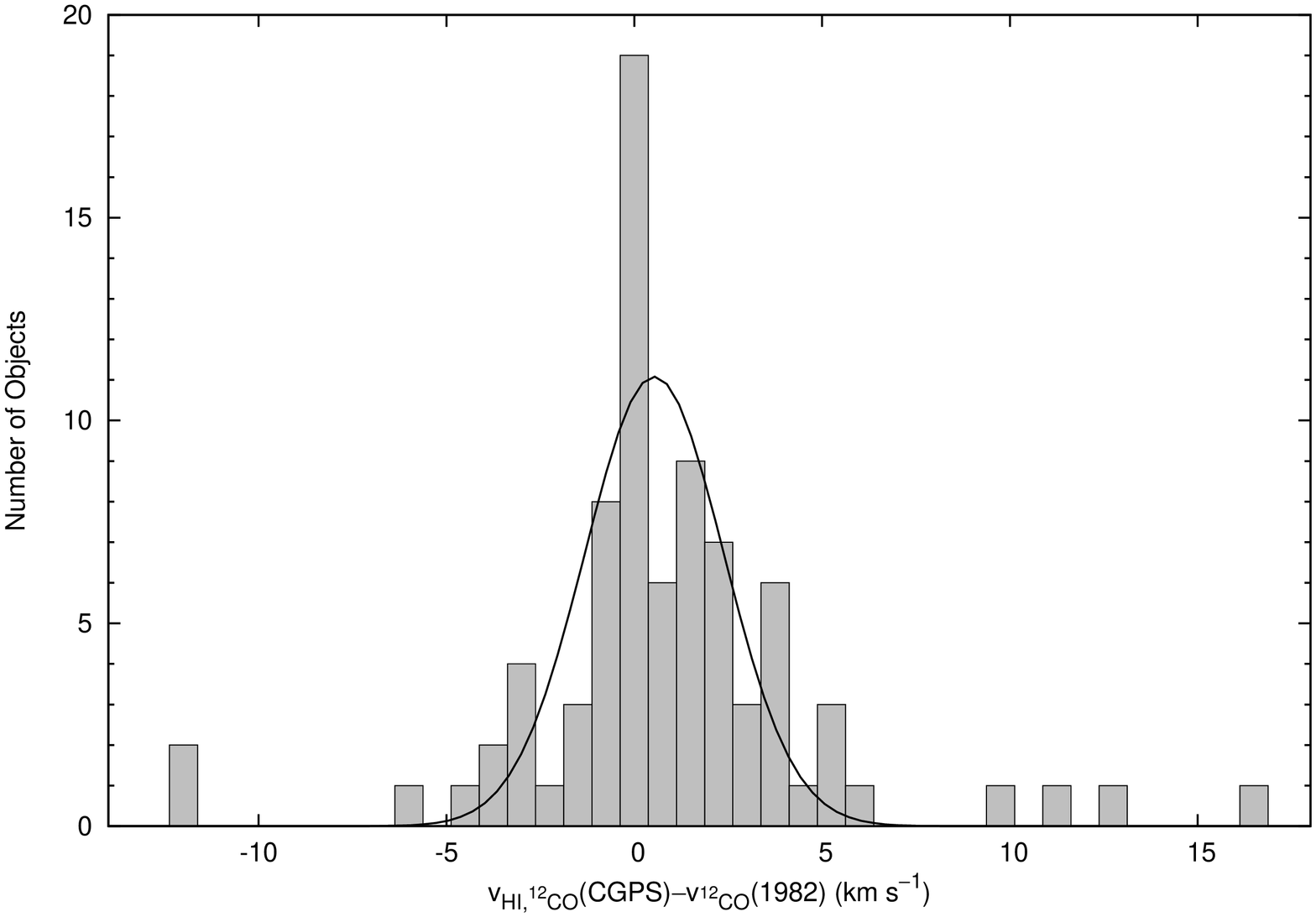}\vspace{-0.5cm}
\end{center}
\caption{(Top) The distribution of differences between $^{12}$CO velocities and 
\ion{H}{1} velocities for all 103 \ion{H}{2} regions of our sample. The mean is 
$-$0.2~km~s$^{-1}$ and the (1$\sigma$) dispersion is 3.0~km~s$^{-1}$. (Bottom)
Distribution of differences between the mean of our \ion{H}{1} and $^{12}$CO 
velocities and $^{12}$CO velocities from \citet{blit82} for 81 matching 
\ion{H}{2} regions. The mean difference is 0.5~km~s$^{-1}$ and the dispersion 
is 1.9~km~s$^{-1}$.}
\label{vdist}
\end{figure}

The distribution of CGPS velocities $v_{^{12}\textrm{CO}}-v_{\textrm{HI}}$ 
(Figure~\ref{vdist}) shows a mean difference of $-$0.2~km~s$^{-1}$ and a 
1$\sigma$ dispersion of 3.0~km~s$^{-1}$, demonstrating good correlation between 
the different gaseous tracers. Comparison of $^{12}$CO velocities with those of 
81 matched \ion{H}{2} regions in \citet{blit82} shows a mean difference of 
0.6~km~s$^{-1}$ and 1$\sigma=~$1.4~km~s$^{-1}$, and between \ion{H}{1} (with 
its broader thermal linewidth) and the $^{12}$CO of \citet{blit82} shows a mean 
of 1.1~km~s$^{-1}$ and 1$\sigma$ width of $\pm~$3.6~km~s$^{-1}$. This indicates 
that small-magnitude extra motions of order $\sim$2-3~km~s$^{-1}$ exist in the 
$^{12}$CO velocities of \citet{blit82} that are absent from the ones here, 
probably the kind demonstrated in Figure~\ref{sh173} (outflows and expansions).
Because of the presence of these random LOS motions, the mean of $^{12}$CO and 
\ion{H}{1} velocities should have somewhat less uncertainty. Indeed, the 
distribution of differences between CGPS mean CO+\ion{H}{1} velocities and
the CO velocities in \citeauthor{blit82} (bottom panel, Fig.~\ref{vdist}) shows 
a mean of 0.5~km~s$^{-1}$ and 1$\sigma$ variation of $\pm~$1.9~km~s$^{-1}$.

This dispersion in the differences of CO and \ion{H}{1} velocities 
($\pm~$3.0~km~s$^{-1}$) is taken as the uncertainty in the systemic velocity for 
each object in each tracer. This should include variations from random thermal 
and cloud-cloud motions in each tracer, instrumental linewidths, and typical 
variations due to the estimation by eye itself, probably the dominant source 
here. We also recognize that some individual \ion{H}{2} regions have less 
reliable systemic velocities due to multiple velocity components identified in 
each tracer, so $\pm~$3.0~km~s$^{-1}$ is the \textit{minimum} uncertainty in
each velocity estimate.


\begin{longtable*}{ccccccccc}



\tabletypesize{\scriptsize}


\tablecaption{}

\tablenum{2}

\tablehead{\colhead{Sh2-no.} & \colhead{$\ell$} & \colhead{$b$} &
\colhead{$v_{^{12}\textrm{CO}}$} & \colhead{$v_{\textrm{HI}}$} &
\colhead{$r$} & \colhead{$dr=\sigma_*/\sqrt{n_*}$} &
\colhead{$n_{*}$ stars} & \colhead{$E\left(B-V\right)$(mag)} \\ 
\colhead{/Name} & \colhead{($\degr$)} & \colhead{($\degr$)} &
\colhead{($\pm~$3.0~km~s$^{-1}$)} & \colhead{($\pm~$3.0~km~s$^{-1}$)} &
\colhead{(kpc)} & \colhead{(kpc)} &
\colhead{(\# excl.)} & \colhead{mean of $n_{*}$} } 

\startdata
121 & 90.23 & +1.72 & $-$62.4 & $-$56.7 & 6.82 & 0.32 & 3 & 2.022 \\
124 & 94.57 & $-$1.45 & $-$43.55 & $-$41.87 & 3.78 & 0.64 & 2 & 1.319 \\
CTB104b & 94.72 & $-$1.54 & $-$14.95 & $-$14.65 & 2.36 & 0.14 & 5 & 0.638 \\
127 & 96.27 & +2.57 & $-$94.7 & $-$92.98 & 9.97 & 1.99 & 1 & 1.73 \\
BFS8 & 96.35 & $-$0.2 & $-$54.44 & $-$55.47 & 8.75 & 1.75 & 1 & 1.338 \\
128 & 97.56 & +3.16 & $-$74 & $-$70.72 & 8.06 & 1.61 & 1 & 1.768 \\
129 & 99.06 & +7.4 & +1.42 & $-$6.3 & 0.81 & 0.15 & 5 & 0.572 \\
G99.1+7.4 & 99.1 & +7.51 & $-$12.6 & $-$8.9 & 1.29 & 0.26 & 1 & 0.598 \\
131 & 99.43 & +3.66 & $-$1.88 & +1.94 & 1.00 & 0.08 & 12 & 0.500 \\
BFS10 & 101.44 & +2.66 & $-$61.3 & $-$60.82 & 6.18 & 1.24 & 1 & 1.73 \\
DA568 & 101.1 & +2.5 & $-$64.5 & $-$63.3 & 4.87 & 0.04 & 2 & 1.084 \\
132 & 102.96 & $-$0.8 & $-$49.28 & $-$49.28 & 3.44 & 0.29 & 6 & 0.825 \\
134 & 103.72 & +2.18 & $-$16.3 & $-$17.54 & 1.63 & 0.33 & 1(1) & 1.038 \\
135 & 104.59 & +1.37 & $-$20.84 & $-$17.13 & 1.4 & 0.28 & 1 & 0.955 \\
137 & 105.63 & +8 & $-$8.08 & $-$11.15 & 0.81 & 0.17 & 7 & 0.27 \\
138 & 105.62 & +0.34 & $-$49.9 & $-$56.7 & 3.04 & 0.61 & 1 & 2.415 \\
139 & 105.77 & $-$0.1 & $-$48.87 & $-$52.99 & 3.22 & 0.29 & 3(1) & 0.61 \\
140 & 106.79 & +5.31 & $-$8.08 & $-$7.24 & 0.92 & 0.16 & 4 & 0.823 \\
141 & 106.81 & +3.31 & $-$65.77 & $-$62.89 & 9.92 & 1.98 & 1 & 1.259 \\
142 & 107.28 & $-$0.9 & $-$39.19 & $-$39.39 & 3.48 & 0.24 & 9 & 0.628 \\
143 & 107.21 & $-$1.34 & $-$33.21 & $-$34.03 & 3.84 & 0.77 & 1 & 0.655 \\
145 & 108.18 & +5.15 & $-$7.01 & $-$7.44 & 0.93 & 0.12 & 2 & 1.241 \\
147 & 108.26 & $-$1.07 & $-$55.05 & $-$51.35 & 3.13 & 0.63 & 1 & 1.128 \\
148 & 108.36 & $-$1.06 & $-$55.47 & $-$51.76 & 3.14 & 0.63 & 1 & 1.24 \\
149 & 108.39 & $-$1.05 & $-$55.47 & $-$51.76 & 5.46 & 1.09 & 1(4) & 0.83 \\
150 & 108.86 & +6.15 & $-$7.94 & $-$10.53 & 1.02 & 0.12 & 2 & 0.6 \\
151 & 108.6 & $-$2.74 & $-$53.82 & $-$50.11 & 3.26 & 0.26 & 9 & 0.71 \\
152 & 108.76 & $-$0.95 & $-$51.76 & $-$49.28 & 2.9 & 0.58 & 1 & 1.291 \\
153 & 108.77 & $-$0.99 & $-$51.76 & $-$50.11 & 4.6 & 0.41 & 2 & 0.748 \\
154 & 108.98 & +1.59 & $-$8.88 & $-$17.95 & 1.48 & 0.3 & 2 & 1.349 \\
155 & 110.22 & +2.55 & $-$8.88 & $-$9.71 & 0.84 & 0.04 & 10 & 0.825 \\
156 & 110.11 & +0.05 & $-$50.9 & $-$48.46 & 2.68 & 0.54 & 1 & 1.278 \\
BFS17 & 110.20 & +0.01 & $-$50.11 & $-$47.22 & 2.73 & 0.55 & 1 & 0.855 \\
157 & 111.2 & $-$0.75 & $-$47.3 & $-$47.43 & 2.93 & 0.28 & 9 & 0.8 \\
158 & 111.54 & +0.78 & $-$50.93 & $-$50.93 & 2.44 & 0.77 & 2 & 1.614 \\
159 & 111.61 & +0.37 & $-$54.23 & $-$60 & 3.07 & 0.61 & 1 & 1.232 \\
160 & 111.8 & +3.73 & $-$10.53 & $-$15.07 & 0.95 & 0.09 & 3 & 0.749 \\
161 & 112.1 & +1.02 & $-$47.63 & $-$45.57 & 2.96 & 0.59 & 1 & 1.2 \\
162 & 112.23 & +0.24 & $-$45.16 & $-$52.17 & 2.41 & 0.16 & 2(2) & 0.51 \\
163 & 113.52 & $-$0.57 & $-$43.93 & $-$41.86 & 3.01 & 0.41 & 5(1) & 1.132 \\
164 & 113.91 & $-$1.6 & $-$37.95 & $-$43.51 & 3.08 & 0.62 & 1 & 1.02 \\
165 & 114.65 & +0.14 & $-$34.44 & $-$35.27 & 1.96 & 0.39 & 1 & 0.76 \\
166 & 114.5 & $-$0.86 & $-$48.05 & $-$50.52 & 2.36 & 0.47 & 1(2) & 0.988 \\
168 & 115.79 & $-$1.65 & $-$41.04 & $-$38.57 & 2.14 & 0.30 & 5(1) & 0.865 \\
169 & 115.88 & $-$1.71 & $-$37.74 & $-$38.98 & 2.09 & 0.42 & 1 & 0.84 \\
170 & 117.57 & +2.26 & $-$46.81 & $-$48.05 & 2.79 & 0.33 & 6(1) & 0.649 \\
171 & 118.2 & +4.99 & $-$14.65 & $-$15.89 & 0.91 & 0.09 & 7 & 1.388 \\
173 & 119.4 & $-$0.84 & $-$32.38 & $-$29.49 & 2.96 & 0.31 & 10(4) & 0.53 \\
175 & 120.36 & +1.97 & $-$49.28 & $-$51.76 & 2.67 & 0.53 & 1(2) & 1.052 \\
177 & 120.75 & $-$0.28 & $-$47.22 & $-$45.98 & 2.37 & 0.22 & 6(1) & 0.508 \\
180 & 122.63 & +0.09 & $-$43.51 & $-$40.63 & 5.41 & 1.08 & 1 & 0.77 \\
G122.6+1.6 & 122.67 & +1.45 & $-$55.88 & $-$48.05 & 2.41 & 0.15 & 5 & 0.7 \\
184 & 123.15 & $-$6.29 & $-$30.4 & $-$27.02 & 3.10 & 0.13 & 8(1) & 0.368 \\
185 & 123.96 & $-$1.8 & $-$31.69 & $-$31.97 & 3.23 & 0.65 & 1 & 0.48 \\
186 & 124.9 & +0.1 & $-$41.86 & $-$42.69 & 2.76 & 0.15 & 2 & 1.07 \\
187 & 126.67 & $-$0.805 & $-$15.48 & $-$15.48 & 1.58 & 0.32 & 1(1) & 1.431 \\
G127.1+0.9 & 127 & +0.84 & $-$47.43 & $-$46.8 & 2.15 & 0.43 & 1 & 0.62 \\
190 & 133.71 & +1.21 & $-$42.69 & $-$42.28 & 2.00 & 0.13 & 8(1) & 0.825 \\
192 & 136.13 & +2.08 & $-$46.81 & $-$50.11 & 3.49 & 0.7 & 1 & 0.946 \\
193 & 136.14 & +2.12 & $-$45.98 & $-$51.76 & 2.44 & 0.49 & 1(1) & 0.76 \\
196 & 136.44 & +2.54 & $-$45.16 & $-$47.63 & 5.52 & 1.1 & 1 & 0.835 \\
G137.8$-$1.0 & 137.77 & $-$0.95 & $-$103.7 & $-$102.46 & 6.93 & 1.39 & 1 & 1.021 \\
198 & 137.38 & +0.2 & $-$44.75 & $-$40.01 & 2.49 & 0.5 & 1(2) & 0.95 \\
199 & 137.7 & +1.6 & $-$37.75 & $-$34.03 & 1.76 & 0.14 & 8(1) & 0.645 \\
LBN676 & 139.66 & +2.54 & $-$42.69 & $-$43.51 & 3.5 & 0.7 & 1 & 0.95 \\
202 & 139.99 & +2.09 & $-$10.33 & $-$14.66 & 0.97 & 0.08 & 5(1) & 0.586 \\
G140.8+3.1 & 140.8 & +3.06 & $-$8.88 & $-$10.53 & 0.6 & 0.12 & 1 & 0.708 \\
BFS28 & 141.73 & +2.76 & $-$11.05 & $-$11.77 & 0.73 & 0.12 & 2(1) & 0.499 \\
203 & 143.75 & $-$1.75 & $-$32.38 & $-$33.21 & 1.79 & 0.36 & 1 & 1.391 \\
BFS31 & 143.82 & $-$1.57 & $-$31.88 & $-$30.73 & 1.75 & 0.35 & 1 & 0.73 \\
204 & 145.83 & +2.94 & $-$36.95 & $-$32.79 & 3.76 & 0.14 & 6(2) & 0.653 \\
205 & 148 & $-$0.4 & $-$6.48 & $-$14.24 & 0.85 & 0.15 & 3(2) & 0.607 \\
206 & 150.61 & $-$0.93 & $-$23.75 & $-$21.67 & 3.02 & 0.6 & 1 & 1.356 \\
207 & 151.21 & +2.11 & $-$25.78 & $-$28.88 & 4.27 & 1.19 & 2(1) & 1.06 \\
208/Wat~1 & 151.29 & +1.97 & $-$30.1 & $-$25.79 & 4.44 & 0.55 & 4(3) & 0.89 \\
209 & 151.61 & $-$0.24 & $-$48.87 & $-$50.11 & 10.58 & 0.57 & 3 & 1.676 \\
211 & 154.65 & +2.46 & $-$37.97 & $-$37.33 & 7.39 & 0.44 & 3 & 1.664 \\
212 & 155.36 & +2.615 & $-$34.41 & $-$37.12 & 4.81 & 0.6 & 4(1) & 0.867 \\
217 & 159.15 & +3.27 & $-$20.43 & $-$20.84 & 4.37 & 0.87 & 1 & 0.725 \\
219 & 159.36 & +2.57 & $-$25.37 & $-$23.72 & 4.23 & 0.64 & 2(1) & 0.849 \\
223 & 166.2 & +2.54 & $-$22.49 & $-$20.02 & 4.05 & 0.81 & 1 & 0.66 \\
225 & 168.09 & +3.07 & $-$22.49 & $-$23.73 & 3.79 & 0.76 & 1 & 0.721 \\
227 & 168.68 & +1.09 & $-$19.33 & $-$17.54 & 4.08 & 0.82 & 1(3) & 0.831 \\
228 & 169.19 & $-$0.9 & $-$13.01 & $-$14.65 & 4.2 & 0.3 & 2 & 1.4 \\
229 & 172 & $-$2.2 & $-$3.11 & +4.72 & 0.73 & 0.24 & 2 & 0.491 \\
231 & 173.47 & +2.55 & $-$18.92 & $-$18.16 & 2.12 & 0.42 & 1 & 1.131 \\
232 & 173.5 & +3.1 & $-$12.83 & $-$11.63 & 2.09 & 0.42 & 1(1) & 0.565 \\
234 & 173.38 & $-$0.19 & $-$8.51 & $-$12.8 & 2.19 & 0.1 & 13 & 0.54 \\
235 & 173.62 & +2.81 & $-$17.4 & $-$16.72 & 1.36 & 0.27 & 1 & 1.18 \\
236 & 173.6 & $-$1.78 & $-$3.81 & $-$20.22 & 4.03 & 0.32 & 7(1) & 0.54 \\
237 & 173.97 & +0.25 & $-$4.7 & $-$12.46 & 3.76 & 0.28 & 3 & 0.723 \\
241 & 180.79 & +4.03 & $-$6.41 & $-$12.19 & 4.84 & 0.97 & 1 & 0.621 \\
242 & 182.36 & +0.19 & +1.42 & +2.65 & 2.19 & 0.44 & 1 & 0.74 \\
247 & 188.96 & +0.85 & +2.65 & +1.83 & 2.23 & 0.18 & 2 & 0.978 \\
249 & 189.45 & +4.38 & +0.18 & +8.84 & 2.01 & 0.15 & 5 & 0.555 \\
252 & 189.81 & +0.33 & +8.02 & +12.55 & 2.23 & 0.14 & 11 & 0.633 \\
253 & 192.23 & +3.59 & +14.63 & +12.55 & 4.29 & 0.37 & 8(1) & 0.525 \\
254 & 192.44 & $-$0.21 & +12.55 & +14.61 & 2.43 & 0.49 & 1 & 0.645 \\
255 & 192.63 & $-$0.02 & +12.55 & +11.72 & 2.27 & 0.45 & 1 & 1.18 \\
256 & 192.60 & $-$0.13 & +7.6 & +14.61 & 2.59 & 0.52 & 1 & 1.345 \\
257 & 192.58 & $-$0.08 & +7.6 & +11.31 & 2.16 & 0.43 & 1 & 0.865 \\
258 & 192.72 & +0.04 & +8.02 & +15.85 & 3.03 & 0.61 & 1 & 1.398 \\
259 & 192.91 & $-$0.62 & +22.86 & +23.68 & 8.71 & 1.74 & 1 & 1.228 \\
\hline
\multicolumn{9}{l}{} \\
\multicolumn{5}{l}{Additional \ion{H}{2} regions beyond CGPS} & & & & \\

261 & 194.148 & $-$2.037 & - & - & 1.89 & 0.38 & 1 & 0.61 \\
267 & 196.188 & $-$1.177 & - & - & 3.89 & 0.28 & 2 & 1.123 \\
268 & 196.38 & $-$2.85 & 4.8 & - & 3.26 & 0.07 & 3 & 0.697 \\
269 & 196.45 & $-$1.68 & 17.5 & - & 4.27 & 0.85 & 1 & 1.365 \\
270 & 196.83 & $-$3.10 & 25.6 & - & 9.27 & 1.85 & 1 & 1.07 \\
271 & 197.8 & $-$2.33 & 20.5 & - & 3.90 & 0.47 & 1 & 0.98 \\
272 & 196.83 & $-$2.33 & 20.6 & - & 4.96 & 0.99 & 1 & 0.938\enddata






\end{longtable*}

\section{Comparison with Previous Studies}
Like the more recent catalog of \citet{russ03}, we present a homogeneous set of 
new stellar distances to 103 \ion{H}{2} regions derived using a common LC and 
colour calibration for stars. Our approach is more fundamental, however, in 
that we use high-resolution radio maps to define the outermost boundary of the 
\ion{H}{2} emission and then consider (or reconsider) all known OB stars in the 
area, excluding those outside of the emission bounds or outside a distance 
range that is common to the others. This data mining has greatly increased the 
number of individual SFRs with consistently calculated stellar distance 
estimates in quadrants II and part of III: for example, the compilation of 
\citet{hou09} lists only 56 single or group distances for 71 objects in the 
longitude range of the CGPS. 

\begin{figure}
\begin{center}
\vspace{-0.5cm}
\includegraphics[angle=-90,width=9cm]{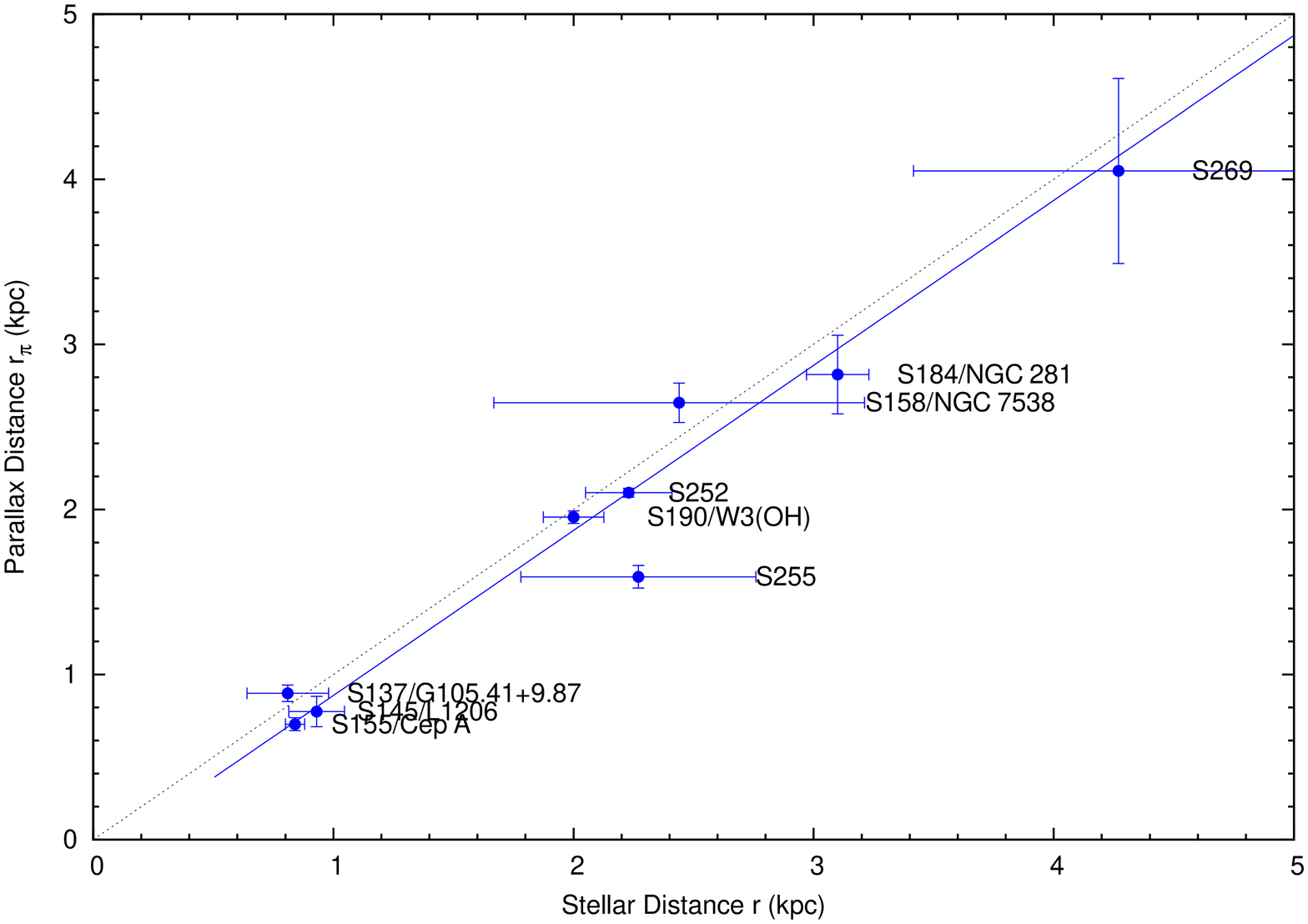}
\includegraphics[angle=-90,width=9cm]{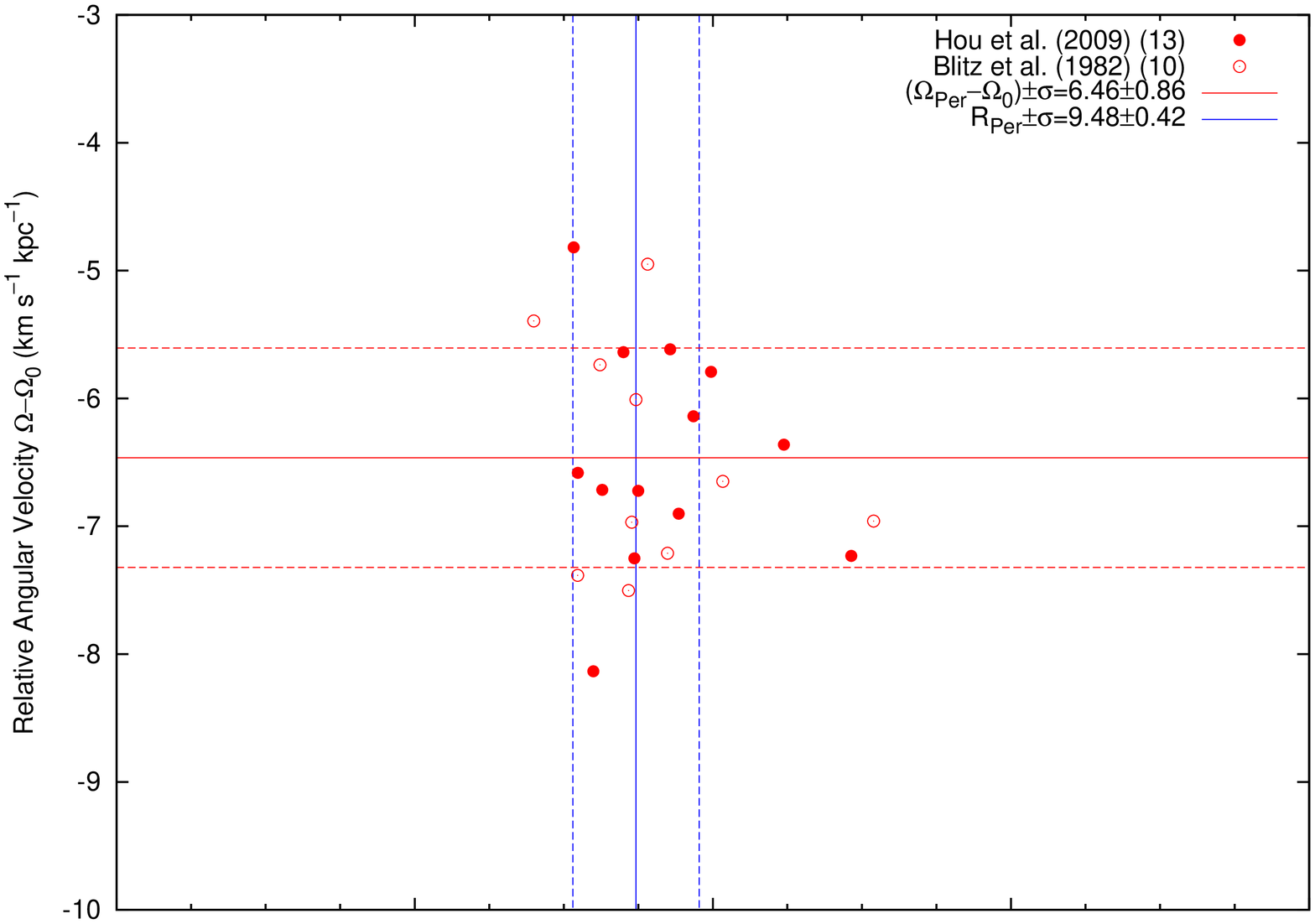}\vspace{-0.42cm}
\includegraphics[angle=-90,width=9cm]{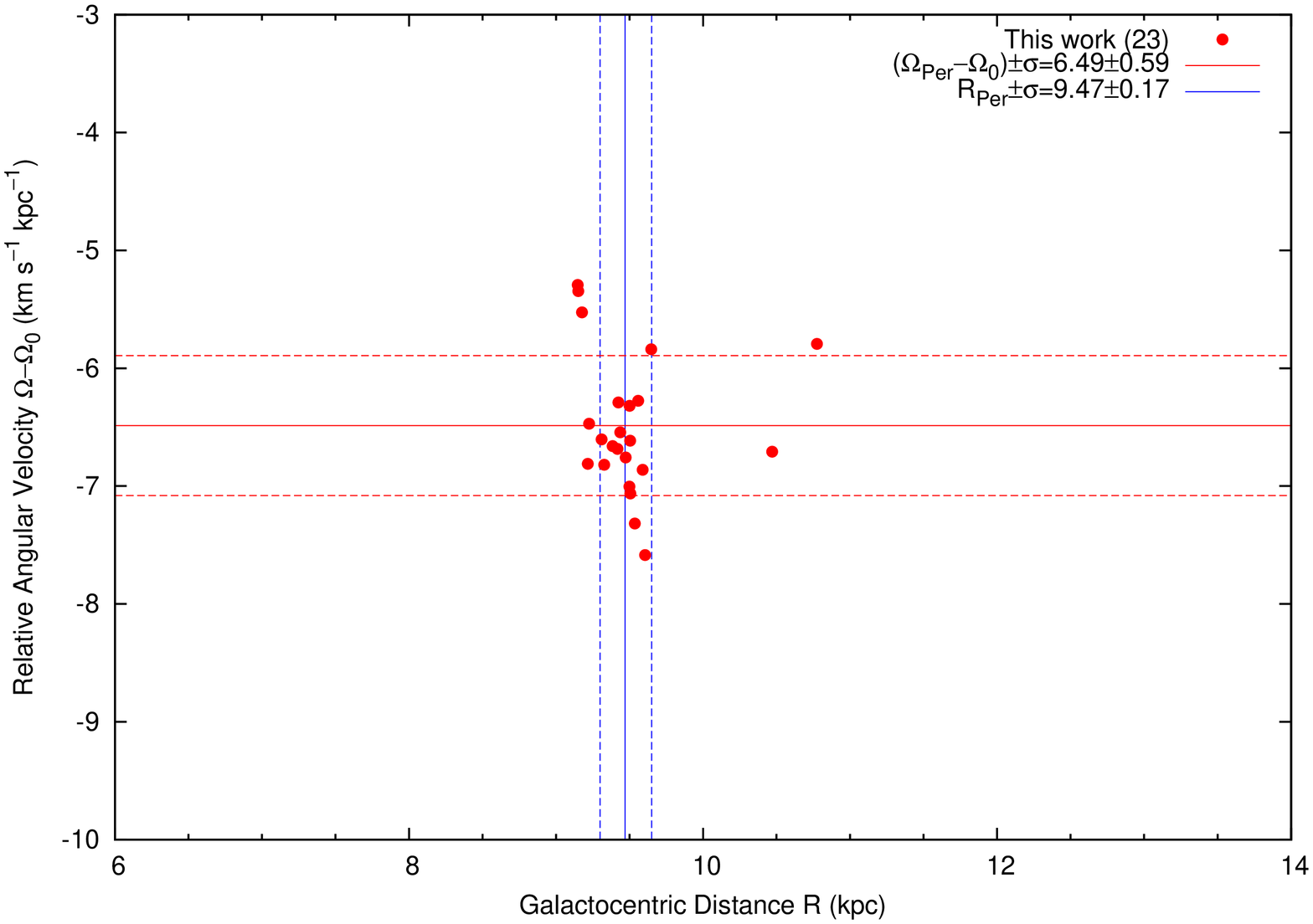}\vspace{-0.5cm}
\end{center}
\caption{(Top) Comparison of our stellar distances to parallax distances 
$r_{\pi}$ for 9 \ion{H}{2} regions in the CGPS matched with masers observed 
with VLBI. References are \citet{reid09} (Cep~A, NGC~7538, NGC~281, W3(OH), 
Sh2-252), \citet{rygl10} (L1206 and Sh2-255), \citet{xu13} 
(Sh2-137/G105.41+9.87) and \citet{asak14} (Sh2-269). (Centre and Bottom) 
Comparison of Galactocentric distances and angular velocities for 23 Perseus 
spiral arm \ion{H}{2} regions, calculated with heliocentric distances and LSR 
velocities from \citet{hou09} and \citet{blit82} (centre) and from Table~2 here 
(bottom). Dashed lines indicate the 1$\sigma$ (standard deviation) of the 
sample in each axis. See Sec.~\ref{improve}.} 
\label{vlbi}
\end{figure}

Nine high-mass SFRs with distances and uncertainties measured via VLBI 
observations of trigonometric parallax can be matched to \ion{H}{2} regions in 
the CGPS (Figure~\ref{vlbi}; top). The error-weighted correlation between these 
distances ($r_{\pi}\pm dr_{\pi}$) and our stellar distances (plotted in 
Figure~\ref{vlbi}) is $r_{\pi}=~$0.999$\pm$0.054$r$-0.126$\pm$0.070, 
essentially 1:1 within a distance range up to $r\lesssim$4.5~kpc. Although not 
within the CGPS region, we also include our stellar distance to Sh2-269 of 
$r=~$4.3$\pm$20\%~kpc \citep[from the lone B0.5V star found by][see 
{}``Additional \ion{H}{2} regions beyond CGPS'' in Table~2 and online 
Table~1]{moff79} and use the recent parallax results of \citet{asak14}:
$r_{\pi}=~$4.05$^{+0.65}_{-0.49}$~kpc. 

\subsection{Improvements}\label{improve}
Improvements in the distances and velocities of \ion{H}{2} regions will 
translate into a clearer view of Galactic structure and dynamics. These 
improvements can be assessed by measuring the scatter in Galactocentric 
distance $R$ and angular velocity relative to the Sun $\Omega-\Omega_0$ 
\citep[Eqn.~1 in][]{fost10} in spiral arm objects, both values for which are 
expected to be (nearly) constant across a (nearly) circular ring about the 
Galactic centre (assuming the pitch angle of the arm is small and only a short 
segment of the arm is considered). We have 34 Perseus spiral arm \ion{H}{2} 
regions between 90$\degr\lesssim\ell\lesssim~$150$\degr$ \citep[see Paper 
II][for this determination]{fost13}. We compare distances $R_{Per}$ from our 
catalogue to the compilation of \citet{hou09}, which has 19 objects in common 
\citep[velocities and distances mainly referenced 
from][]{pala03,fbs89,russ03,russ07}. Eight additional objects with stellar 
distances and CO velocities are found in \citet{blit82}. A value for the Solar 
galactocentric distance $R_0=~$8~kpc \citep[][]{fost10} is assumed. Some 
objects in \citet{hou09} are {}``grouped'' together into 2 entries with 
duplicated distances and velocities. For proper comparison to our data (which 
is not grouped) we remove one member of the Sh2-148/149 and Sh2-203/BFS~31 
groups from the comparison (149,203), and replace one member in 3 other groups 
(Sh2-138/139, 152/153 and 158/159) with alternate stellar distances and 
velocities from \citet{blit82}. We also remove the distance outliers BFS~8 and 
Sh2-180 from both datasets. 

Two plots of $\Omega-\Omega_0$ vs. $R_{Per}$ for the 23 remaining objects are 
shown in Figure~\ref{vlbi}, one made with previous catalogue data and the other 
from data in Table~2. Visually, the results of the current study cluster 
together much tighter in each variable as compared to previous catalogue 
values. A robust direct comparison of the median and normalized median absolute 
deviations in $R_{Per}$ using distances in \citet{hou09,blit82} show 
$R_{Per}\simeq~$9.48$\pm$0.42~kpc, whereas for the same regions in our data 
$R_{Per}\simeq~$9.47$\pm$0.17~kpc; essentially the same median distance to the 
Perseus arm but with significantly lower relative scatter. 
In the comparison of angular velocities, a mean and 1$\sigma$ scatter of 
$\Omega-\Omega_0~=-$6.455$\pm$0.857~km~s$^{-1}$~kpc$^{-1}$ for the Perseus 
spiral arm is found for the 23 objects catalogued by \citet{hou09} and 
\citet{blit82,fbs89}. Referenced velocities in \citet{hou09} are a mix of CO 
and radio recombination line \citep[][]{pala03}; if we restrict the comparison 
to strictly $^{12}$CO velocities from \citet{blit82,fbs89} the scatter is  
$\pm~$0.809~km~s$^{-1}$~kpc$^{-1}$. Using CGPS $^{12}$CO velocities alone the 
scatter is reduced to $\pm~$0.592~km~s$^{-1}$~kpc$^{-1}$, and with CGPS 
\ion{H}{1} it is $\pm~$0.686~km~s$^{-1}$~kpc$^{-1}$. Taking the mean between 
the two gaseous tracers (weighted by 1/$\sigma^2$) results in 
$-$6.486$\pm~$0.593~s$^{-1}$~kpc$^{-1}$, a factor of $\approx1/\sqrt{2}$ less 
scatter in $\Omega-\Omega_0$ than in the same sample from previous catalogues. 
In particular, using $^{12}$CO velocities in this work also produces
$\sim$1/$\sqrt{2}$ less scatter than using those from \citet{blit82}. Clearly, 
the technique of estimating the systemic velocity from related gas seen edge-on 
results in a dataset better for studying Galactic dynamics, having reduced the 
variations due to random (e.g. {}``turbulence'' and cloud-to-cloud motions) as 
well as expansion and outflow motions along the LOS.

In summary, the new distances and CGPS velocities presented in this paper do 
indeed have favourable characteristics for use in tracing Galactic structure 
and dynamics, and have an edge in terms of lower scatter when compared to 
existing data in the literature.

\acknowledgements
We would like to thank the referees for their careful reading of and thoughtful 
comments on our manuscript. The Dominion Radio Astrophysical Observatory is 
operated as a national facility by the National Research Council of Canada. The 
Canadian Galactic Plane Survey has been a Canadian project with international 
partners, and was supported by grants from the Natural Sciences and Engineering 
Research Council of Canada (NSERC). The Five College Radio Astronomy 
Observatory was supported by NSF grant AST 0540852. TF has been supported by an 
NSERC Discovery grant and a Brandon University Research Committee Grant. CB is 
funded in part by the UK Science and Technology Facilities Council grant 
ST/J001627/1 (``From Molecular Clouds to Exoplanets'') and the ERC grant 
ERC-2011-StG\_20101014 (``LOCALSTAR''), both held at the University of Exeter.

\end{document}